%% file: PPS-17-001_temp.tex
\pdfoutput=1

\documentclass[11pt,twoside,a4paper,cmspaper,final,collab]{cms-tdr}
\def\svnVersion{471675P}\def\cmsCernNoTag{CERN-EP-2018-014}\def\cmsCernDate{\today}\def\cmsMessage{Published in the Journal of High Energy Physics as \href{http://dx.doi.org/10.1007/JHEP07(2018)153}{\doi{10.1007/JHEP07(2018)153.}}}

\begin{document}\cmsNoteHeader{PPS-17-001}

\hyphenation{had-ron-i-za-tion}
\hyphenation{cal-or-i-me-ter}
\hyphenation{de-vices}
\RCS$Revision: 471675 $
\RCS$HeadURL: svn+ssh://alverson@svn.cern.ch/reps/tdr2/papers/PPS-17-001/trunk/PPS-17-001.tex $
\RCS$Id: PPS-17-001.tex 471675 2018-08-11 20:44:25Z alverson $

\newlength\cmsFigWidth
\ifthenelse{\boolean{cms@external}}{\setlength\cmsFigWidth{0.85\columnwidth}}{\setlength\cmsFigWidth{0.4\textwidth}}
\ifthenelse{\boolean{cms@external}}{\providecommand{\cmsLeft}{left\xspace}}{\providecommand{\cmsLeft}{left\xspace}}
\ifthenelse{\boolean{cms@external}}{\providecommand{\cmsRight}{right\xspace}}{\providecommand{\cmsRight}{right\xspace}}

\newlength\cmsTabSkip\setlength{\cmsTabSkip}{1ex}

\renewcommand{\cmsCollabName}{The CMS and TOTEM Collaborations}
\renewcommand{\cmslogo}{\includegraphics[height=2.33cm]{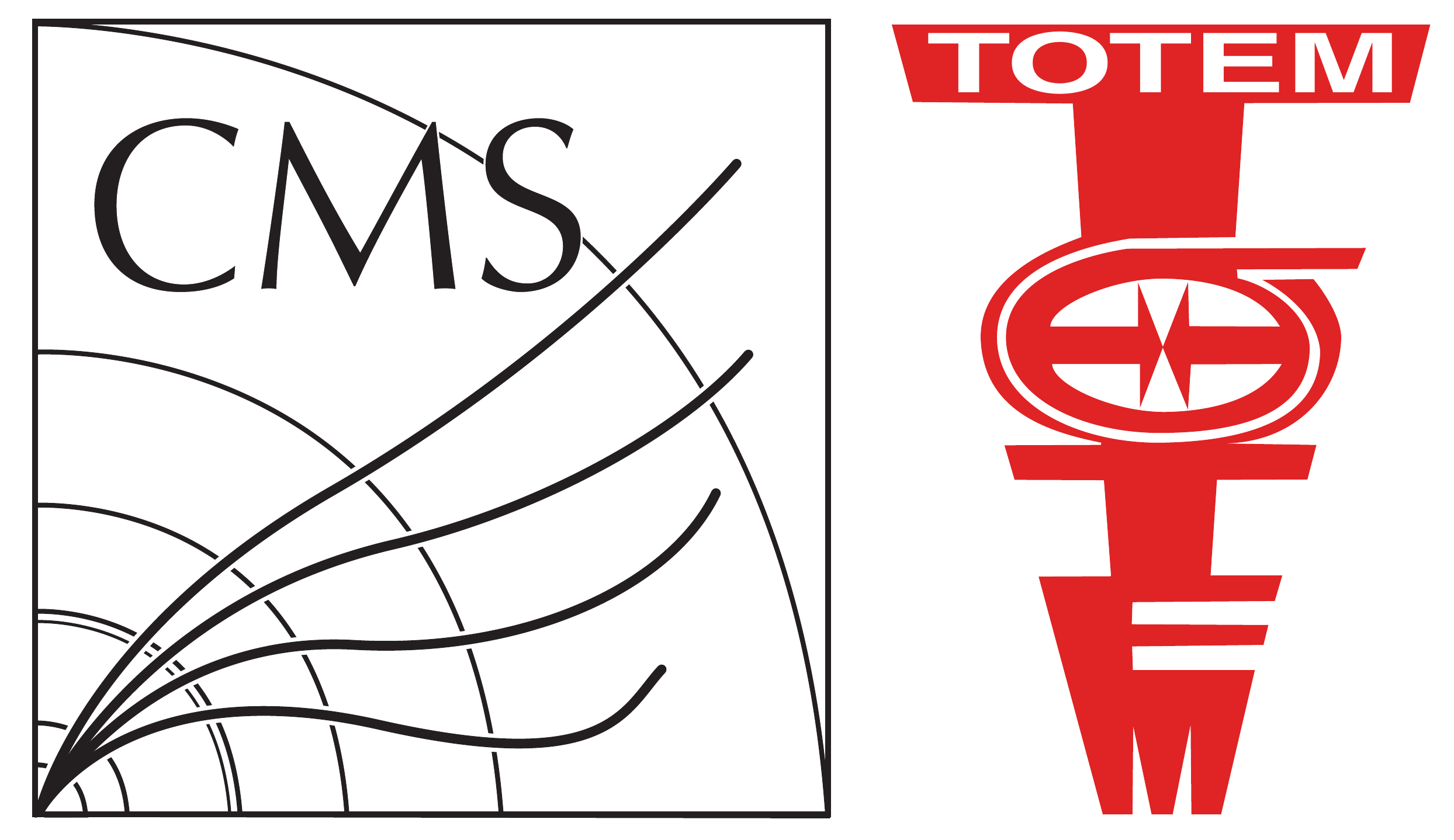}}

\renewcommand{\cmsTag}{CMS-\cmsNUMBER\\&TOTEM 2018-001\\}
\renewcommand{\cmsPubBlock}{\begin{tabular}[t]{@{}r@{}l}&CMS \cmsSTYLE~PPS-17-001\\&TOTEM-NOTE-2017-003\\\end{tabular}}
\renewcommand{\appMsg}{See Appendices~\ref{app:collab} and \ref{app:totem} for the lists of collaboration members}
\renewcommand{\cmsCopyright}{\copyright\,\the\year\ CERN for the benefit of the CMS and TOTEM Collaborations.}

\providecommand{\NA}{\ensuremath{\text{---}}}
\providecommand{\CL}{CL\xspace}

\cmsNoteHeader{PPS-17-001}

\title{Observation of proton-tagged, central (semi)exclusive production of
high-mass lepton pairs in $\Pp\Pp$ collisions at 13\TeV with the CMS--TOTEM precision proton spectrometer}

\date{\today}

\abstract{
The process $\Pp\Pp \to \Pp \ell^+\ell^- \Pp^{(*)}$,
with $\ell^+\ell^-$ a muon or an electron pair produced at midrapidity with mass larger than 110\GeV,
has been observed for the first time at the LHC in $\Pp\Pp$ collisions at $\sqrt{s} = 13\TeV$.
One of the two scattered protons is measured in the CMS--TOTEM precision proton spectrometer (CT--PPS),
which operated for the first time in 2016. The second proton either remains intact or is excited and then dissociates into a low-mass state
$\Pp^{*}$, which is undetected. The measurement is based on an integrated
luminosity of 9.4\fbinv collected during standard,
high-luminosity LHC operation. A total of 12 \MM and 8 \EE
pairs with $m(\ell^{+}\ell^{-}) > 110\GeV$, and matching forward proton
kinematics, are observed, with expected backgrounds of $1.49 \pm 0.07 \stat \pm 0.53 \syst$
and $2.36 \pm 0.09 \stat \pm 0.47 \syst$, respectively. This corresponds to an excess of more
than five standard deviations over the expected background.
The present result constitutes the first observation of proton-tagged $\PGg\PGg$
collisions at the electroweak scale. This measurement also demonstrates that CT--PPS performs according to the design specifications.
}

\hypersetup{
pdfauthor={CMS Collaboration},
pdftitle={Observation of proton-tagged, central (semi)exclusive production
of high-mass lepton pairs in pp collisions at 13 TeV with the CMS-TOTEM precision proton
spectrometer},
pdfsubject={CMS},
pdfkeywords={CMS, physics, photon-photon fusion }}

\maketitle

\section{Introduction}

Proton--proton collisions at the LHC provide for the first time the conditions to study the
production of particles with masses at the electroweak scale via photon--photon fusion~\cite{deFavereaudeJeneret:2009db,dEnterria:2008iks}.
Although the production of high-mass systems in photon--photon collisions has been observed by the
CMS and ATLAS experiments~\cite{Chatrchyan:2013akv, Khachatryan:2016mud,Aaboud:2016dkv}, no such measurement exists so
far with the simultaneous detection of the scattered protons. This paper reports the
measurement of the process $\Pp\Pp \to \Pp \ell^+\ell^- \Pp^{(*)}$ in $\Pp\Pp$ collisions at $\sqrt{s} = 13\TeV$,
where a pair of leptons ($\ell = \Pe$, $\mu$) with mass  $m(\ell^+\ell^-) > 110\GeV$ is reconstructed in the central CMS apparatus, one of the protons is
detected in the CMS--TOTEM precision proton spectrometer (CT--PPS), and the second proton either
 remains intact or is excited and then dissociates into a low-mass state, indicated by the symbol $\Pp^{*}$,
and escapes undetected. Such a final state receives contributions from exclusive, $\Pp\Pp \to \Pp\ell^{+}\ell^{-}\Pp$,
and semiexclusive, $\Pp \Pp \to \Pp \ell^{+} \ell^{-}\Pp^{*}$, processes (Fig. 1 left, and center). Central exclusive dilepton
production is interesting because deviations from the theoretically well-known cross section may be
an indication of new physics~\cite{Atag:2009vz,Sahin:2009gq,Inan:2010af}, whereas central semiexclusive processes constitute a background
to the exclusive reaction when the final-state protons are not measured.

(Semi)exclusive dilepton production has been previously studied at the Fermilab Tevatron and at the
CERN LHC, but at lower masses and never with a proton
tag~\cite{Abulencia:2006nb, Aaltonen:2009cj, Chatrchyan:2011ci, Chatrchyan:2012tv, Aad:2015bwa, Aaboud:2017oiq}.
In this paper, forward protons are reconstructed in CT--PPS, a near-beam magnetic spectrometer that uses
the LHC magnets between the CMS interaction point (IP) and detectors in the TOTEM area about 210 m away on both sides of the IP~\cite{Albrow:1753795}. Protons that have
lost a small fraction of their momentum are bent out of the beam envelope, and their trajectories
are measured.

\begin{figure}[h!]
  \centering
     \includegraphics[width=.9\textwidth]{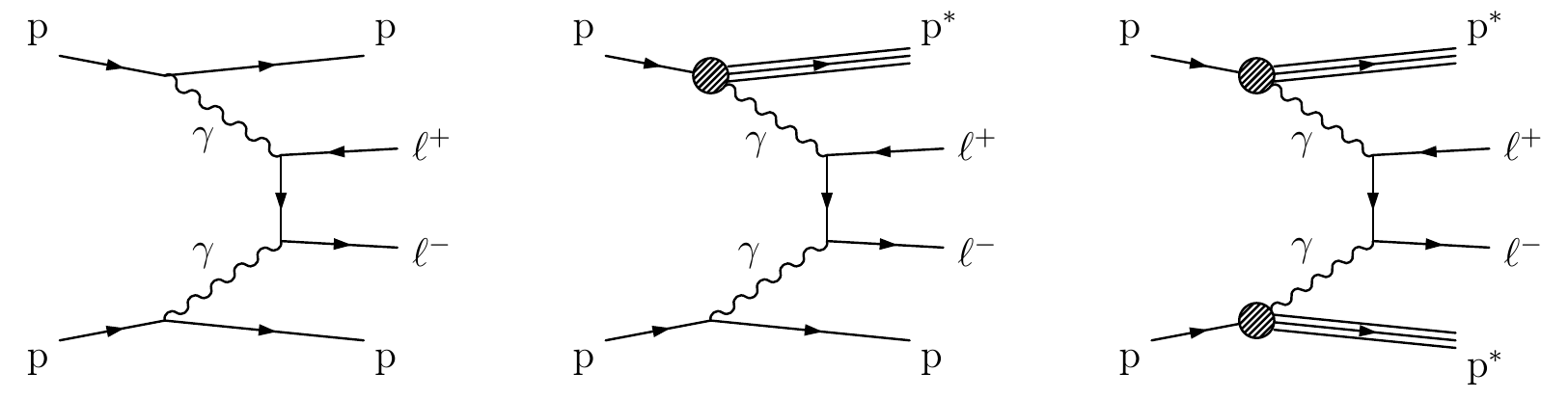}
    \caption{Production of lepton pairs by $\PGg\PGg$ fusion. The exclusive (left), single proton dissociation or
semiexclusive (middle), and double proton
dissociation (right) topologies are shown. The left and middle processes
result
in at least one intact final-state proton, and are considered signal in this
analysis. The rightmost diagram is considered to be a background process.}
  \label{fig:FeynmanDiagrams}
\end{figure}

Central dilepton production is dominated by the diagrams shown in Fig.~\ref{fig:FeynmanDiagrams}, in which both protons radiate quasi-real
photons that interact and produce the two leptons in a $t$-channel process. The left and center
diagrams result in at least one intact final-state proton, and are
considered as signal in this analysis. The CT--PPS acceptance for
detecting both protons in ``exclusive'' $\Pp\Pp \to \Pp\ell^{+}\ell^{-}\Pp$
events (the left diagram) starts only above $m(\ell^+\ell^-) \approx 400\GeV$, where the standard model cross section is
small. By selecting events with only a single tagged proton, the sample contains a mixture
of lower mass exclusive and single-dissociation ($\Pp \Pp \to \Pp \ell^{+} \ell^{-}\Pp^{*}$, ``semiexclusive'') processes
with higher cross sections. The right diagram of Fig.~\ref{fig:FeynmanDiagrams} is
considered background, and contributes if a proton from the diffractive dissociation is
detected, or if a particle detected in CT--PPS from another interaction in the same bunch
crossing (pileup), or from beam-induced background is wrongly associated with the
dilepton system. A pair of leptons from a Drell--Yan process can also mimic a signal event if detected in combination with a pileup proton.

In central (semi)exclusive events, the kinematics of the dilepton system can be used to determine
the momentum of the proton, and hence its fractional momentum loss $\xi$. Comparison of this indirect measurement of $\xi$
 with the direct one obtained with CT--PPS can be used to suppress
backgrounds, as well as to provide proof of the correct functioning of the spectrometer.

The CT--PPS detector~\cite{Albrow:1753795,Ruggiero:2009zz} operated for the first time in 2016 and collected
a total integrated luminosity of $\sim$15\fbinv in standard, high-luminosity runs of the LHC. The average number of pileup
interactions per bunch crossing during 2016 was 27. For the present analysis, a sample of 9.4\fbinv is used; the remaining (unused)
data set was taken after September 2016, when the LHC collided protons with
a different crossing angle.

The paper is organized as follows. Section~\ref{apparatus} describes the
experimental setup, and Sections~\ref{alignment}--\ref{proton_reconstruction}
the procedures to derive the alignment and the LHC optics
parameters from the data. Section~\ref{datasets} documents the samples of
data and simulated events used in the analysis, while
Sections~\ref{EventSelection} and~\ref{backgrounds} explain the event
selection criteria, and the methods applied to estimate the backgrounds,
respectively.  Finally, the analysis and the results are presented in
Section~\ref{results}, followed by a summary in Section~\ref{summary}.

\section{Experimental setup}
\label{apparatus}

The central feature of the CMS apparatus is a superconducting solenoid of
6\unit{m} internal diameter, providing a magnetic field of 3.8\unit{T}.
Within the solenoid volume are a silicon pixel and strip tracker with
coverage in pseudorapidity up to $\abs{\eta} = 2.5$, a lead tungstate crystal electromagnetic
calorimeter, and a brass and scintillator hadron calorimeter,
each composed of a barrel and two endcap sections. Forward
calorimeters extend the coverage provided by the barrel and
endcap detectors up to $\abs{\eta} = 5.2$. Muons are measured in gas-ionization
detectors embedded in the steel flux-return yoke outside the solenoid. A
more detailed description of the CMS detector, together with a definition
of the coordinate system used and the relevant kinematic variables, can be
found in Ref.~\cite{Chatrchyan:2008zzk}.

The CT--PPS detector measures protons scattered at small angles and carrying between about 84 to 97\% of
the incoming beam momentum. These protons remain inside the beam pipe and
their trajectory is measured by a system of position-sensitive detectors
at a distance of about 210\unit{m} from the IP, on both sides of CMS. These tracking detectors
 are complemented by timing counters to measure the proton arrival time. The
detector planes are inserted horizontally into the beam pipe by means of
``Roman Pots" (RPs), \ie movable near-beam devices that allow
the detectors to be brought very close (down to a few mm) to the beam without affecting the
vacuum, beam stability, or other aspects of the accelerator operation.

\begin{figure}[h!]
  \centering
\includegraphics[width=.95\textwidth]{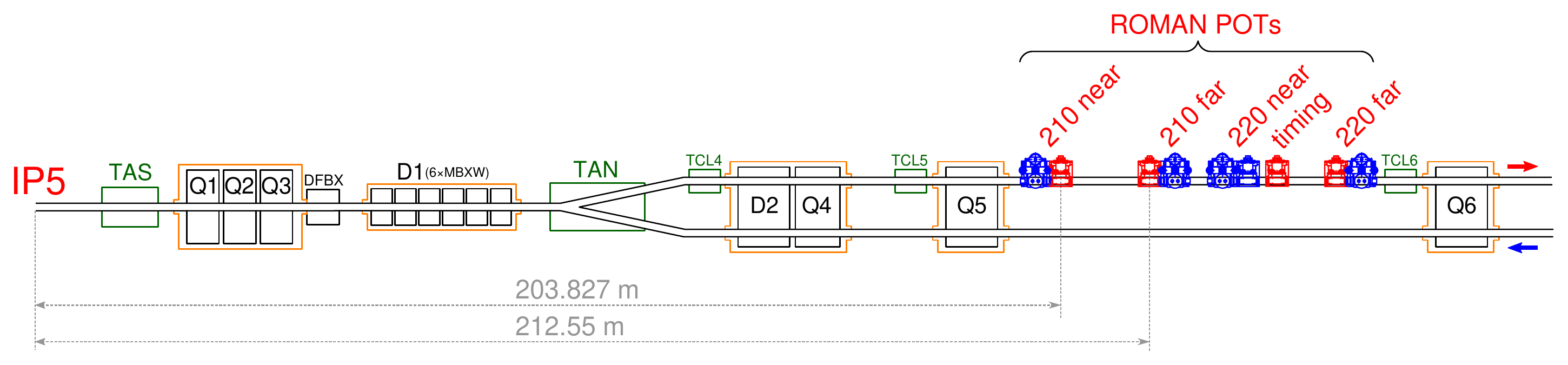}
    \caption{Schematic layout (not to scale) of the beam line as seen
from above between the interaction point
(IP5) and the region where the RPs are located in LHC sector 56. Dipole magnets
(D1, D2) of single- (MBXW) and twin-aperture, quadrupoles (Q1--Q6), collimators (TCL4--TCL6), absorbers (TAS, TAN), and
quadrupole feedboxes (DFBX) are shown. The 210 near and 210 far
units are indicated, along with the timing RPs not used here. The
220 near and 220 far units (not used here) are also shown. The RPs indicated in red are
the horizontal CT--PPS ones; those in blue are part of the TOTEM experiment. The red (blue) arrow indicates the outgoing (incoming)
beam. In the CMS coordinate system, the $z$ axis points to the left. The arm in the opposite
 LHC sector 45 (not shown) is symmetric with respect to the IP.}
  \label{fig:layout}
\end{figure}

\begin{figure}[h!]
  \centering
\includegraphics[width=.55\textwidth]{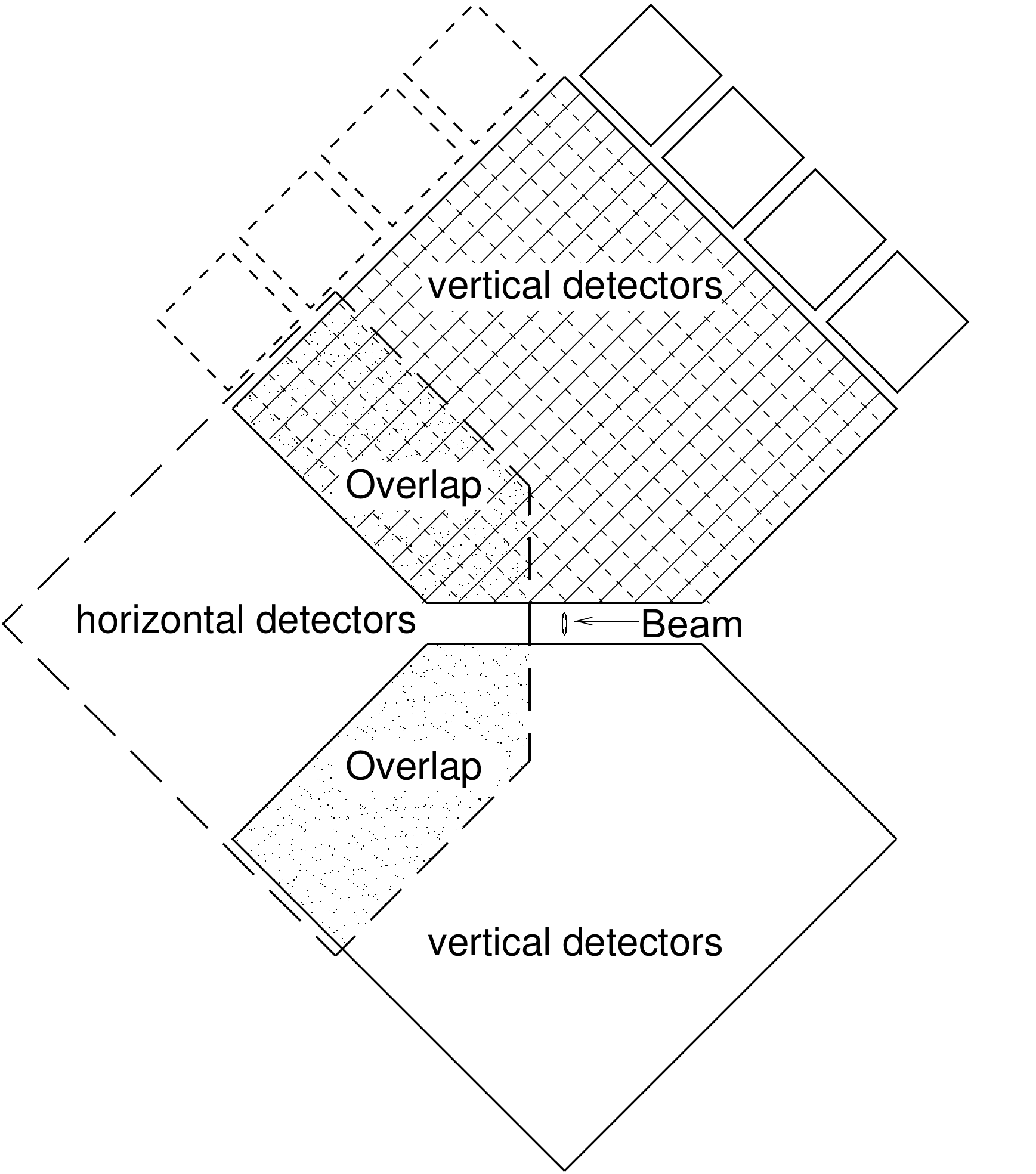}
  \caption{Schematic layout of the silicon strip detectors in one RP station. Both the horizontal RP and the vertical RPs, which are
used only for special low-luminosity calibration fills, are shown. In the top RP, the silicon strips oriented at $+45^{\circ}$ and
$-45^{\circ}$ angles are indicated by the diagonal lines. Tracks in the overlap region, indicated by the shaded area, are used to
perform a relative alignment of the RPs in the calibration fills.}
  \label{fig:sensor}
\end{figure}

The layout of the beam line from the IP to the 210\unit{m} region on one of the
two sides of CMS is shown in Fig.~\ref{fig:layout}. The two sides are
referred to as ``arms" in the following. The arms to the left (positive
$z$ direction) and to the right of CMS when looking from the center of the
LHC correspond to LHC sectors 45 and 56 on the two sides of interaction point 5 where CMS is located, respectively. In each arm
there are two tracking units, referred to as ``210 near" (210N) and ``210
far" (210F), which are located at 203.8\unit{m} and 212.6\unit{m} from the IP, respectively. In 2016,
the tracking RPs were each instrumented with 10 planes of edgeless silicon strip
sensors, providing a spatial resolution of about 12\micron.
Five of these planes are oriented with the silicon strips at a $+45^{\circ}$ angle with
respect to the bottom of the RP, while the other five have the strips at
a $-45^{\circ}$  angle. In total each RP silicon detector plane contains 512 individual strips, with a pitch of 66\micron.
A schematic diagram of the silicon strip sensors, indicating their orientation relative to the LHC beam, is shown
in Fig.~\ref{fig:sensor}

The hit efficiency per plane is estimated to be $>$97\% before the effect of radiation damage to
the sensors. The signal from the silicon detectors is contained within one 25\unit{ns}
bunch crossing of the LHC. The data are read out using a digital VFAT
chip~\cite{Kaplon:2005ce}, and recorded through the standard CMS data acquisition system.
The pots as well as the sensors have been extensively used by the TOTEM experiment and are
described in Refs.~\cite{Antchev:2013hya,Ruggiero:2009zz}. The TOTEM silicon strip sensors were not
designed to sustain exposure to the high radiation doses of the standard high-luminosity
LHC fills. As expected, a first set of such planes suffered severe radiation
damage after about 10\fbinv, and was replaced by a set of spares.
In order to operate at high instantaneous luminosity, the RPs have been equipped
with special ferrite shielding, so as to reduce their electromagnetic impedance, and
hence limit their impact on the LHC beams. The
timing detectors are housed in low-impedance, cylindrical RPs
specially built for CT--PPS, located at 215.7\unit{m} from the IP. They were
equipped with diamond detectors for the last part of the run to complement
the tracking silicon strip detectors. They are not used for the analysis discussed here.
In its final configuration, CT--PPS will use 3D silicon pixel sensors for
tracking and diamond sensors for timing.

The data analyzed for this paper were collected with the RPs at a distance of about 15 $\sigma$ from the beam,
where $\sigma$ is the standard deviation of the spatial distribution of the beam in the transverse direction
pointing to the RP; the values of $\sigma$ range from 0.245\unit{mm} for the 210N RP to 0.14\unit{mm} for the timing RP.

\section{Alignment of the CT--PPS tracking detectors}
\label{alignment}

Alignment of CT--PPS is required in order to determine the
position of the sensors with respect to each other inside a RP, the
relative position of the RPs, and the overall position of the
spectrometer with respect to the beam. An overview of the procedure is
given here; more details are available in Ref.~\cite{Kaspar:2256296}.

The alignment procedure consists of two conceptually distinct parts:

\begin{itemize}

\item Alignment in a special, low-luminosity calibration fill (``alignment
fill"), where RPs are inserted very close to the beam (about $5\,\sigma$).

\item Transfer of the alignment information to the standard,
high-luminosity physics fills.

\end{itemize}

\subsection{Alignment fill}
The first step is the beam-based alignment, the purpose of which is to establish the position of the
RPs with respect to the LHC collimators and the beam. It takes place only once per LHC optics setting.
In this procedure, the TOTEM vertical RPs~\cite{Antchev:2013hya} (cf. Fig.~\ref{fig:sensor}) are used
together with the horizontal CT--PPS RPs. The beam is first scraped
with the collimators so that it develops a sharp edge. Then each RP is moved in small (approximately
10\micron) steps until it is in contact with the edge of the beam, which generates a rapid increase in
the rate observed in the beam-loss monitors close to CT--PPS. At this point, each RP
is at the same distance (in units of $\sigma$) as the collimator, \ie the RP is at the edge of the shadow cast by the
collimator. The necessity to get very close to the beam stems from the need of having the TOTEM
vertical and the CT--PPS RPs overlap.
Data are then taken in this configuration, with the horizontal and vertical RPs at $8$ and $5\,\sigma$, respectively.

The second step consists of determining the relative position of all the sensors in each arm using the
data from the alignment fill. This is achieved by minimizing the residuals between hit positions and
fitted tracks. The track reconstruction is described in Section 4.2.
The position (shift perpendicular to the beam) and rotation (about the beam axis) of each
sensor are thereby determined. While no event selection is necessary (since the method assumes that the tracks
 are linear, which is the case as there are no magnets at the RP location), the most valuable events are those with
tracks reconstructed when the RP detectors overlap, which allow the relative alignment of the RPs to be determined.
The method is applied to several data subsamples in order to verify the stability of the results.

Finally, the alignment of CT--PPS with respect to the beam is performed, again with data from the
dedicated fill. A sample of several thousand elastic scattering events, $\Pp\Pp \to \Pp\Pp$, is used for that purpose. The LHC
optics causes the elastic hit distribution in any vertical RP to have an elliptical shape centered on the beam
position. This symmetry is exploited to determine the position of the RP with respect to the beam.

The uncertainties in the results of the procedure just discussed are 5\unit{mrad} for rotations,
50\micron for horizontal shifts, and 75\micron for vertical shifts.

\subsection{Physics fills}

Since the RPs move, and the beam position can change,
the position of CT--PPS with respect to the beam needs to be redetermined
for each fill. The physics fills are characterized by high intensity with
only the horizontal RPs inserted at much larger distances (about
$15\,\sigma$) from the beam than in the alignment fill, and therefore a different
procedure is employed.

\begin{figure}[!h]
\center
\includegraphics[width=0.6\textwidth]{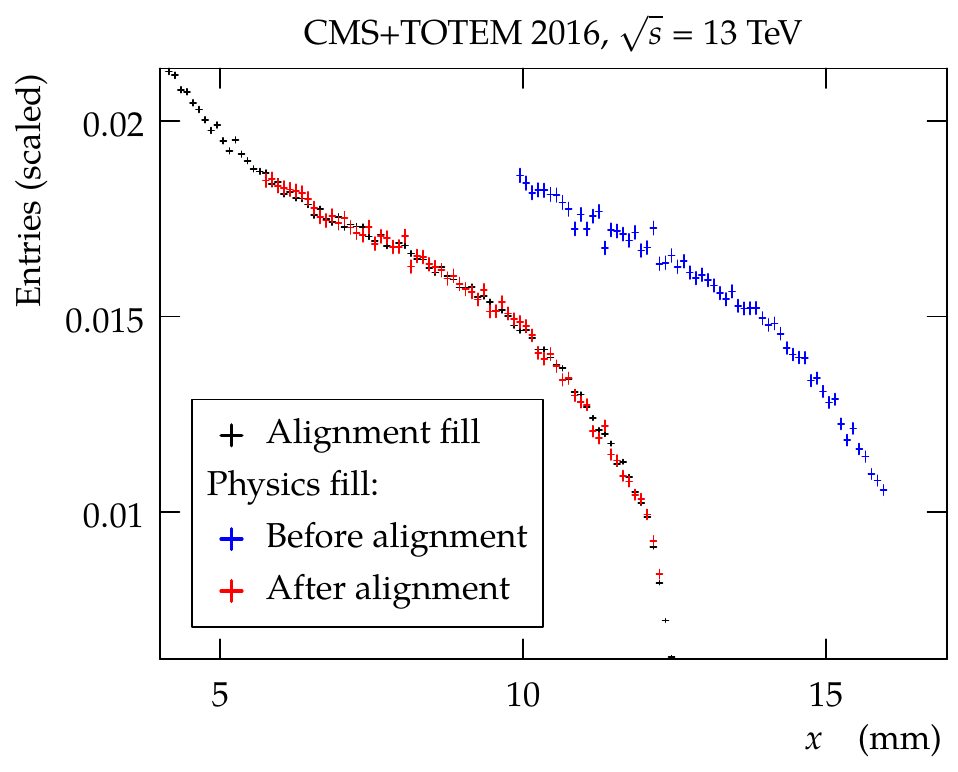}
\caption{Distribution of the track impact points as a function of the
horizontal coordinate for the alignment fill (black points), a physics
fill before alignment (blue points), and after alignment (red points). The
beam center is at $x = 0$ for the black and red points; the $x$ axis origin
is undefined for the blue points. In the alignment procedure the overall normalization of the histogram is irrelevant;
 the histograms from different fills are therefore rescaled to compare their shapes.}
\label{fig-horizontal-match}
\end{figure}

The horizontal alignment is based on the assumption that the
scattered protons from a $\Pp\Pp$ collision at the IP have the same kinematic
distributions in all fills. Given the stability of the LHC conditions (RP positions, collimator setting, magnet currents, and beam orbit),
this leads to the spatial distributions of the track impact points observed in the RPs (Section~4.2).
The alignment is then achieved by matching these distributions
from a physics fill to those from the alignment fill. An example of this
procedure is shown in Fig.~\ref{fig-horizontal-match}. For this method to
work, it is important to suppress the background due to secondary
interactions taking place between the IP and the RPs. To this end, the
correlation between the coordinates of the horizontal hit positions in the
near and far RPs is used. The total uncertainty of the horizontal alignment is
about 150\micron.

The beam vertical position with respect to the sensors is determined by
fitting a straight line to the $y$ coordinate of the maximum of the track
impact point distribution as a function of $x$ (horizontal beam position). The fitted function is then
extrapolated to $x = 0$. This procedure can be
applied since, unlike for the horizontal case, the maximum of the vertical
distribution is within the acceptance of the horizontal RPs. Here again,
the resulting uncertainty is of the order of 150\micron.

\section{Proton reconstruction}
\label{proton_reconstruction}

\subsection{The LHC beam optics}
\label{optics}

\begin{figure}[!htb]
\center
\includegraphics[width=0.8\textwidth]{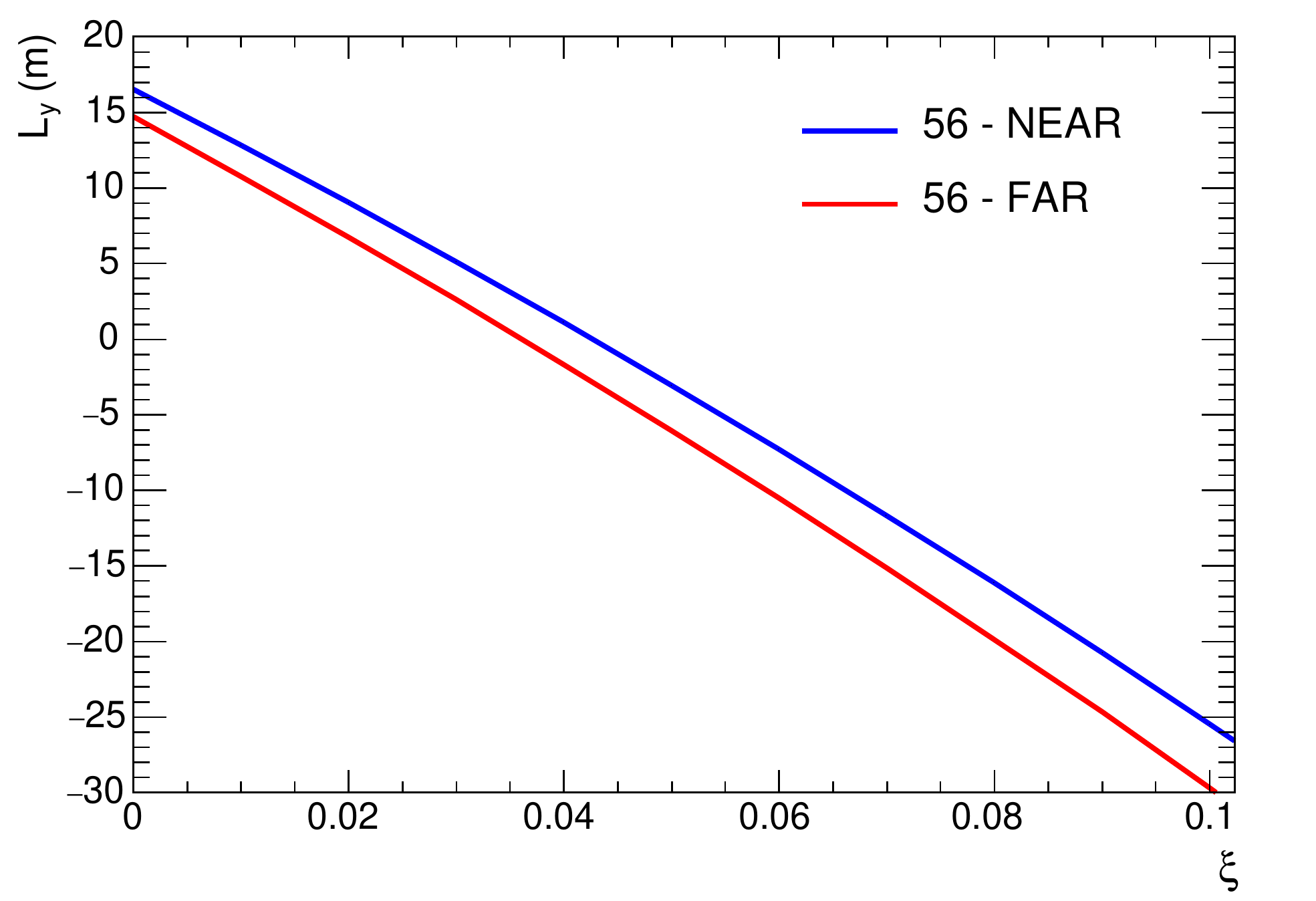}
\caption{Vertical effective length $L_{y}$ (in meters) as a function of the
proton relative momentum loss $\xi$ at two (near and far) RPs calculated
with the beam line optics simulation program~\textsc{mad-x}~\cite{Grote:2003ct}.}
\label{Ly_dependence1}
\end{figure}

\begin{figure}[!h]
\center
\includegraphics[width=0.8\textwidth]{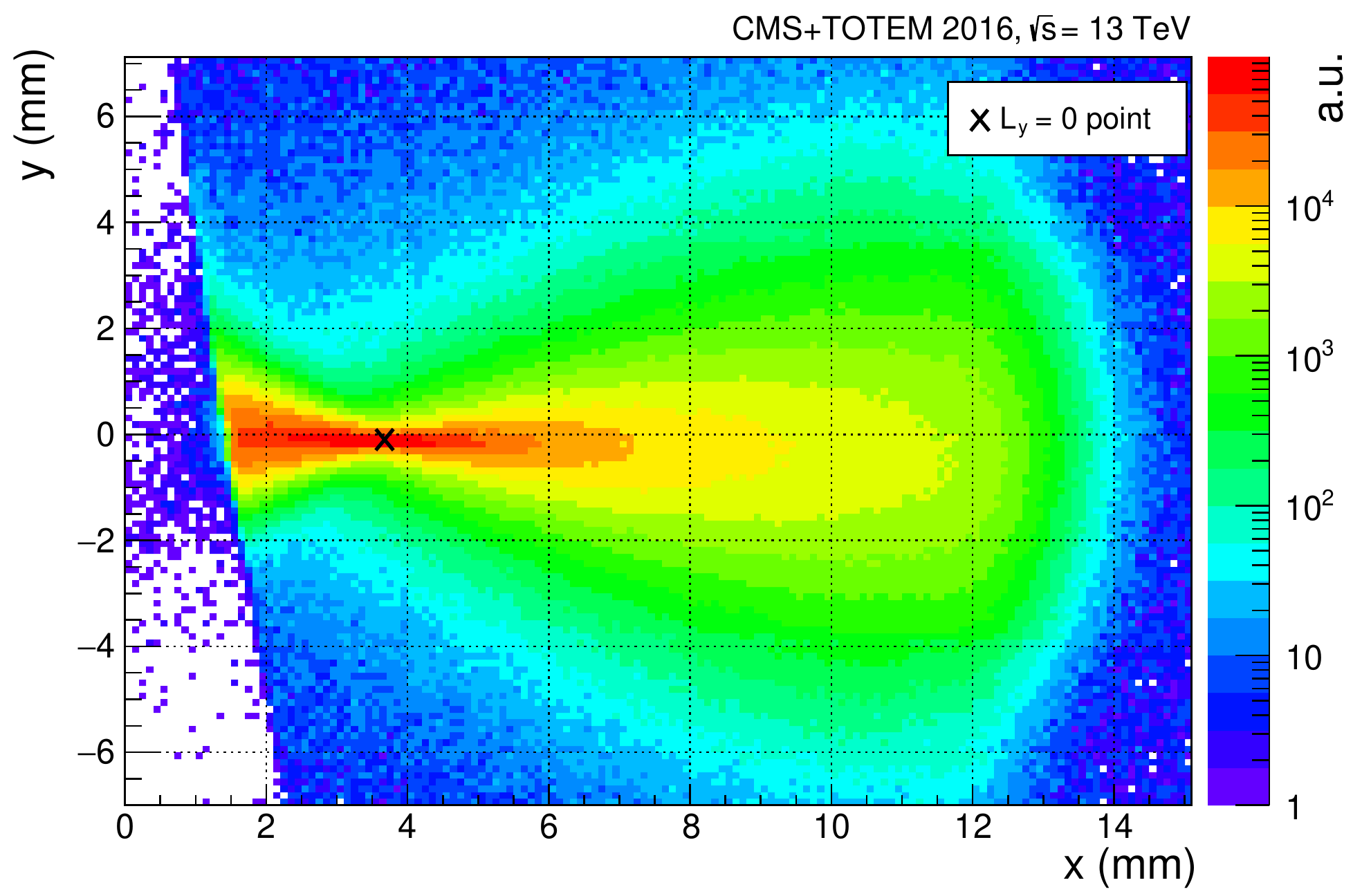}
\caption{
Distribution of the track impact points measured in RP 210F, in sector 45,
for the alignment fill. The point where $L_{y} = 0$ is shown with a cross. The beam center
is at $x = y = 0$. The edge of the distribution is slanted because the RP shown has a rotation of $8^{\circ}$ with respect to the vertical.}
\label{Ly_dependence2}
\end{figure}

The reconstruction of the scattered proton momentum from the tracks measured in the RPs requires
precise knowledge of the magnetic fields traversed by the proton from the IP to the
RPs~\cite{Nemes:2256433}. This is normally parametrized in terms of the
``beam optics", in which the elements of the beam line are
treated as optical lenses. The proton trajectory is described by means of
transport matrices, which transform the kinematics of protons scattered at the IP to
the kinematics measured at the RP position.
The trajectory of protons produced at the IP (denoted by the superscript '$^*$') with transverse vertex position $(x^*,y^*)$ and horizontal and vertical components of scattering angle $(\Theta_x^*, \Theta_y^*)$ is described approximately by:
\begin{equation}
  \mathbf{d}(s)=T(s,\xi)\mathbf{d}^{*},
\end{equation}
\noindent where $s$ indicates the distance from the IP along the nominal beam orbit, and $\mathbf{d} = \left(x,\Theta_x,y,\Theta_y,\xi\right)$, with $\xi = \Delta{p}/p$, and $p$ and $\Delta p$ the nominal beam momentum and the proton longitudinal momentum loss, respectively.
The symbol $T(s,\xi)$ denotes the single-pass transport matrix, whose elements are the optical functions.
The leading term in the horizontal plane is:
\begin{equation}
x = D_{x}(\xi) \xi,
\label{shift_due_to_dispersion}
\end{equation}
where the dispersion $D_{x}$ has a mild dependence on $\xi$.
In the vertical plane, the leading term reads:
\begin{equation}
  y = L_{y}(\xi)\Theta_{y}^{*},
\label{shift_due_to_dispersion2}
\end{equation}
\noindent where $L_{y}(\xi)$ is the vertical effective length.
The $\xi$ dependence of $L_{y}$ is shown in
Fig.~\ref{Ly_dependence1}. At any location $s$ in the RP region there is
a value of $\xi$, $\xi_{0}$, where $L_{y}$ vanishes and hence the values
of $y$
concentrate around zero. Consequently, the distribution of the track
impact points exhibits a `pinch' at $x_{0}\approx D_{x} \xi_{0}\,$, cf.
Fig.~\ref{Ly_dependence2}. The horizontal dispersion $D_{x}$ is then
estimated as:
\begin{equation}
  D_{x} \approx \frac{x_{0}}{\xi_{0}}.
\end{equation}

The subleading terms neglected in this approximation are treated as systematic
uncertainties.

An independent estimate of the difference of the dispersions in the
two LHC beams, $\Delta D_{x}$, is obtained by varying $\Delta D_{x}$
to find the best match between the $\xi$  distributions reconstructed from
the two arms. This estimate agrees with the one discussed above within the uncertainties.

These two horizontal dispersion measurements and the beam position values
constrain the LHC optics between the IP and the RPs,
including the nonlinearities of the proton transport matrices and
their dependence on $\xi$. The optical functions are extracted with the
methods originally developed for the analysis of elastic scattering
data~\cite{Antchev:2014voa}.

\subsection{Proton track reconstruction}

\begin{figure}[h!tb]
  \centering
\includegraphics[width=.65\textwidth]{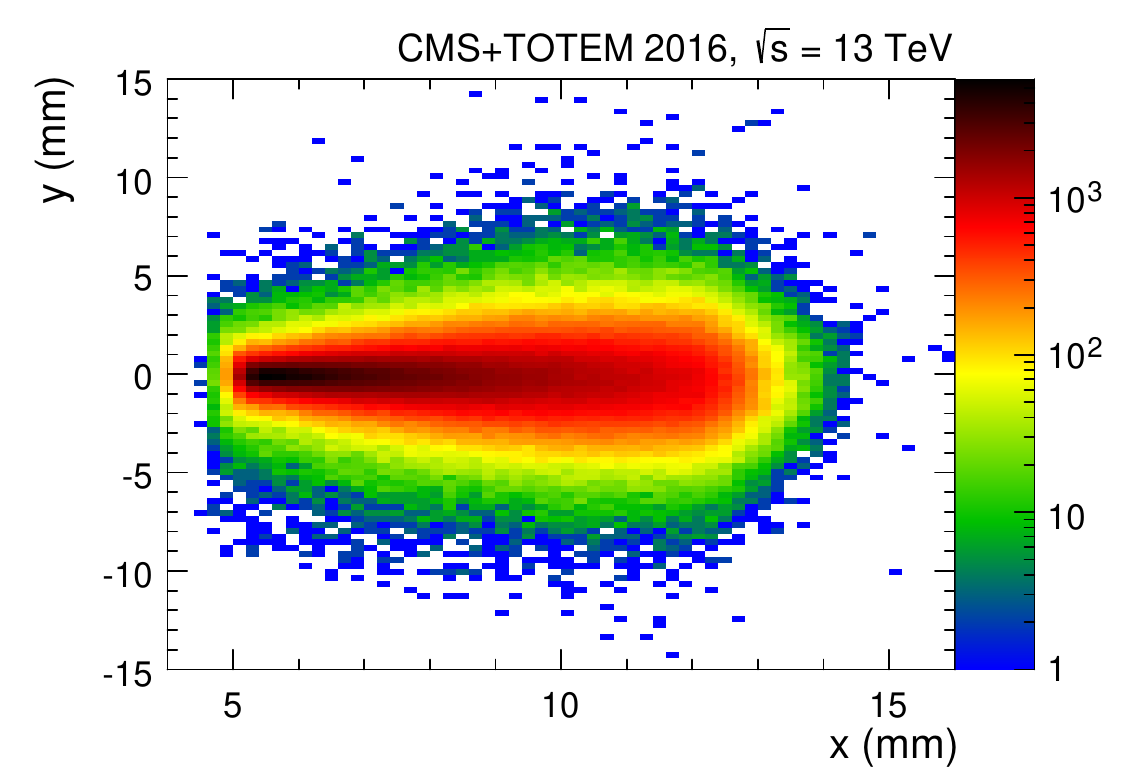}
  \caption{Example of track impact point distribution (in the $x, y$ plane) measured in RP 210F,
sector 45, at $15\,\sigma$ from the beam in the $x$ direction. The beam center is at
$x = y = 0$. The track selection includes a matching requirement with RP 210N, which suppresses noise and beam backgrounds, but slightly reduces the acceptance
for low values of the position $x$, given the different acceptance of the near and far RPs.}
  \label{fig:hit_distribution}
\end{figure}

Since there is no significant magnetic field in the region of the CT--PPS RPs, the trajectory of
particles passing through the silicon strip detectors is a straight line. In each RP (RP hereinafter refers to the particle detector contained in the pot),
 track reconstruction therefore starts with a search for linear patterns along $z$ among the hits detected in the 10 planes, as described in Chapter 3 of Ref.~\cite{Kaspar:2011eva}.
The search is performed independently in each of the two strip orientations (with angles of $+45^{\circ}$
and $-45^{\circ}$ with respect to the bottom of the RP); hits in at least 3 out of 5 planes are required. If only
one pattern is found in both orientations, the patterns can be uniquely associated and a track
fitted, yielding a ``track impact point" evaluated at the center of the RP
along $z$. Figure~\ref{fig:hit_distribution} shows a typical distribution of the track impact
points in the $(x, y)$ plane for a RP at $15\,\sigma$ from the beam. When there is more than one pattern in any strip orientation, a
unique association is not possible and no track is reconstructed.
The inefficiency due to multiple tracks depends on the pileup, and ranges
between 15 and 40\% in the 2016 data used in this analysis. This multiple tracks inefficiency
and the $\xi$- and time-dependent effects of radiation damage to the sensors described in Section 2
are the dominant sources of inefficiency. Other reconstruction effects, such as those due to showers within the
detector material, are estimated to contribute ${\approx}3\%$ to the efficiency for finding proton tracks.

\subsection{Determination of \texorpdfstring{$\xi$}{xi}}

The fractional momentum loss of a proton, $\xi$, can be determined from
the track impact point in a single RP. This is advantageous in regions
where the other RP of the sector does not have sufficient acceptance or is
inefficient. In practice, $\xi$ is reconstructed by inverting
Eq.~(\ref{shift_due_to_dispersion}). This method ignores subleading terms
in the proton transport (notably the one proportional to the horizontal
scattering angle); their effect is included in the systematic
uncertainties.

The main uncertainties are:
\begin{itemize}
\item dispersion calibration: relative uncertainty in $D_x$ of about 5.5\%;
\item horizontal alignment: approximately 150\unit{\mum};
\item neglected terms in Eq.~(\ref{shift_due_to_dispersion}).
\end{itemize}

For values of $\xi \gtrsim 0.04$, the leading uncertainty comes from the dispersion,
and from the neglected terms related to $\Theta^{*}_{x}$ in Eq.~(2).

Having reconstructed $\xi$, Eq.~(\ref{shift_due_to_dispersion2}) can then be
used to determine the vertical scattering angle from the curves presented
in Fig.~\ref{Ly_dependence1}. The scattering angle can, in turn, be related
to the vertical component of the proton transverse momentum.

\section{Data sets and Monte Carlo samples}
\label{datasets}

The CT--PPS data analyzed here were collected during the period May--September 2016;
they correspond to an integrated luminosity of 9.4\fbinv. In the
same period, CMS collected a total of 15.6\fbinv. For the present
data, the beam amplitude function $\beta^*$ at the IP was $0.4$\unit{m} and the crossing angle $\alpha_x$ of the
beams was 370\unit{$\mu$rad}. After about a month, the silicon strip detectors
suffered heavy radiation damage. After new silicon strip detectors were installed in
September, the LHC implemented a smaller crossing angle for collisions at
the CMS IP, which resulted in different optics parameters and therefore changed the CT--PPS acceptance.
These later data are therefore not used in the present analysis.

Simulated signal samples of exclusive ($\Pp\Pp \to \Pp \ell^{+} \ell^{-} \Pp$) and single
proton dissociative ($\Pp\Pp \to \Pp \ell^{+}\ell^{-} \Pp^{*}$) events proceeding via photon fusion
$\PGg \PGg \to \ell^{+}\ell^{-}$ are generated with the \textsc{lpair}
code~\cite{Vermaseren:1982cz,Baranov:1991yq} (version 4.2).
\textsc{lpair} is also used to produce $\PGg\PGg \to \ell^{+} \ell^{-}$
samples with both protons exciting and dissociating, that is $\Pp\Pp \to \Pp^{*} \ell^{+} \ell^{-} \Pp^{*}$.
These three topologies are illustrated in Fig.~\ref{fig:FeynmanDiagrams}.
The central detector information is passed through the standard
\GEANTfour~\cite{Agostinelli:2002hh} simulation of the CMS detector and reconstructed in
the same way as the collision data. Conversely, only generator-level forward proton information is
used, which is sufficient for the present analysis.

Background from the Drell--Yan process, $\Pp\Pp \to \gamma^{*} / \Z^{*} \to \ell^+ \ell^- +X$, is simulated with \newline
\MGvATNLO~\cite{Alwall:2014hca,Frederix:2012ps},
interfaced with the \PYTHIA 8.212~\cite{Sjostrand:2007gs} event generator using the CUETP8M1 tune~\cite{Khachatryan:2015pea}
for parton showering, underlying event, and hadronization. The events are generated at leading order, and normalized to the next-to-next-to-leading order cross
section prediction~\cite{Li:2012wna}.

\section{Event selection}
\label{EventSelection}

\subsection{Central variables}

Events were selected online~\cite{Khachatryan:2016bia} by requiring the presence of at least two muon (electron)
candidates of any charge, each with transverse momentum $\pt > 38\,(33)\GeV$. No requirement on
forward protons was imposed online.

Offline, the tracks of the two highest-$\pt$ lepton candidates of the same flavor in the
event are fitted to a common vertex. The vertex position from the fit is
required to be consistent with that of a collision  ($\abs{z} < 15\unit{cm}$),
with a $\chi^{2} < 10$ (probability greater than 0.16\% for 1 degree of freedom). The lepton
candidates are further required to have $\pt > 50\GeV$, and to pass the standard CMS quality
criteria~\cite{Chatrchyan:2012xi,Khachatryan:2015hwa}. In the final stage of the analysis
only leptons with opposite charge are retained. No explicit isolation is required for the leptons; however, nonprompt
leptons (\ie from heavy and light hadron decays in flight) are heavily suppressed by the applied track multiplicity criteria described below.

\begin{figure}[h!tb]
  \centering
 \includegraphics[width=.99\textwidth]{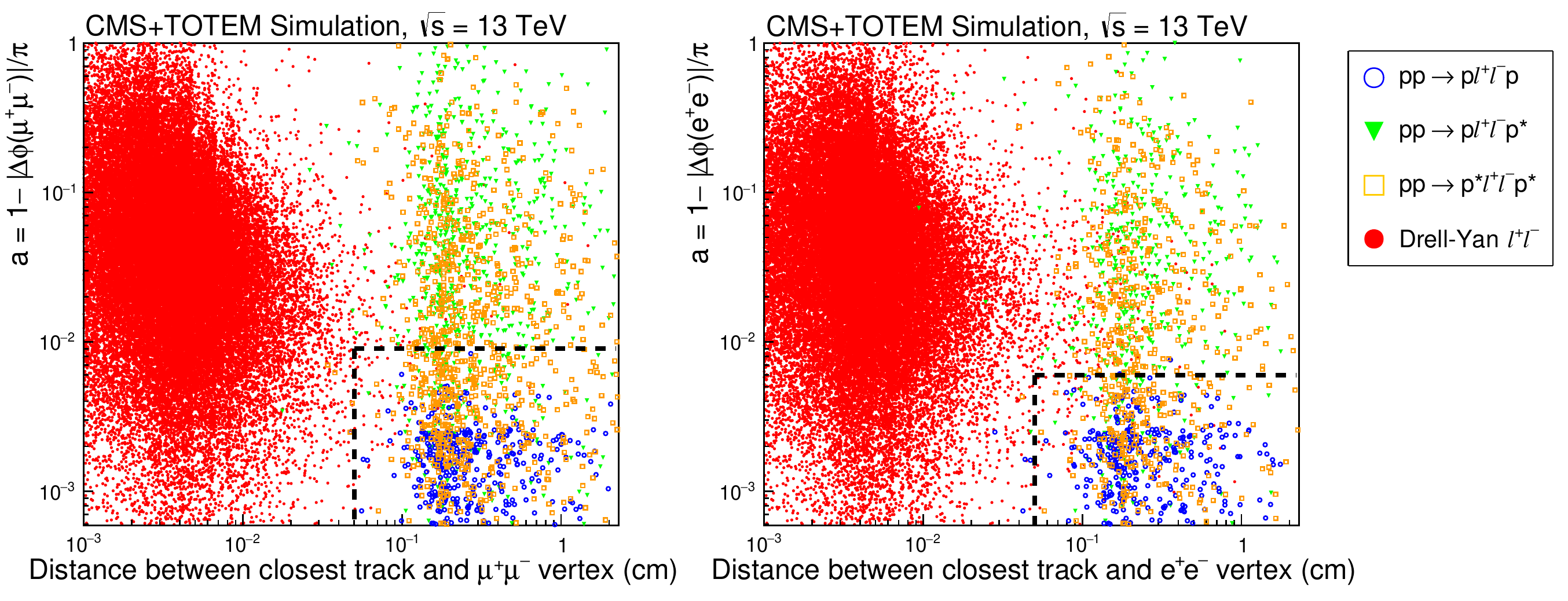}
  \caption{
Dimuon (left) and dielectron (right) acoplanarity versus the distance between the closest
extra track and the dilepton vertex for simulated signal and backgrounds.
The points represent the Drell--Yan (red), exclusive $\PGg\PGg \to \ell^{+}\ell^{-}$
(blue), single-dissociative $\PGg\PGg \to \ell^{+}\ell^{-}$ (green), and double-dissociative
$\PGg\PGg \to \ell^{+}\ell^{-}$ (yellow)
processes. The dashed lines indicate the region selected for the
analysis. The number of points for each physics process does not reflect
its cross section.}
  \label{fig:ntracks_vs_acop_MC}
\end{figure}

In order to select a sample enriched in $\PGg\PGg \to \ell^+\ell^-$ events,
a procedure similar to that of the Tevatron and Run 1 LHC
analyses~\cite{Khachatryan:2016mud, Chatrchyan:2013akv, Chatrchyan:2011ci,
Aaboud:2016dkv, Aad:2015bwa, Abulencia:2006nb, Aaltonen:2009cj,
Aaltonen:2009kg} is used.
The event is accepted if no additional tracks are found in the region within the veto
distance around the dilepton vertex. No explicit requirement is
made on the $\pt$ or on the quality of these extra tracks. In addition, the
dilepton acoplanarity ($a = 1 - \abs{\Delta \phi(\ell^+\ell^-)}/\pi$) is required to be
consistent with the two leptons being back-to-back in azimuth $\phi$.
The dilepton acoplanarity versus the distance between the closest extra track and the
dilepton vertex is shown in Fig.~\ref{fig:ntracks_vs_acop_MC} for muons (left) and electrons (right), for the simulated signal
(blue and green dots) and double-dissociation and Drell--Yan backgrounds (red and yellow dots).
 Based on these distributions, an extra-track veto region distance of at least 0.5\unit{mm} around the vertex is required,
along with $a < 0.009$ for the muons and $a < 0.006$ for the electrons.
The acoplanarity requirements are chosen such that the signal to background
ratio predicted by the simulation is above unity before any matching of
the leptons with RP tracks. The size of the extra-track veto region is smaller than suggested by the
simulation, reflecting the fact that the distribution of primary
vertices in $z$ is narrower in the data than in the simulation. Because
of the high pileup rate, the selection is based on information from reconstructed
tracks alone, without using information from the calorimeters. This results in an efficiency
of $>95\%$ for the highest values of pileup and pileup density observed in the 2016 data set used for
the measurement.

Finally, the invariant mass of the leptons is required to satisfy $m(\ell^+\ell^-) >
110\GeV$. This suppresses the region around the $\Z$ boson mass, which is expected to be
dominated by Drell--Yan production.

\begin{figure}[h!]
  \centering
    \includegraphics[width=.48\textwidth]{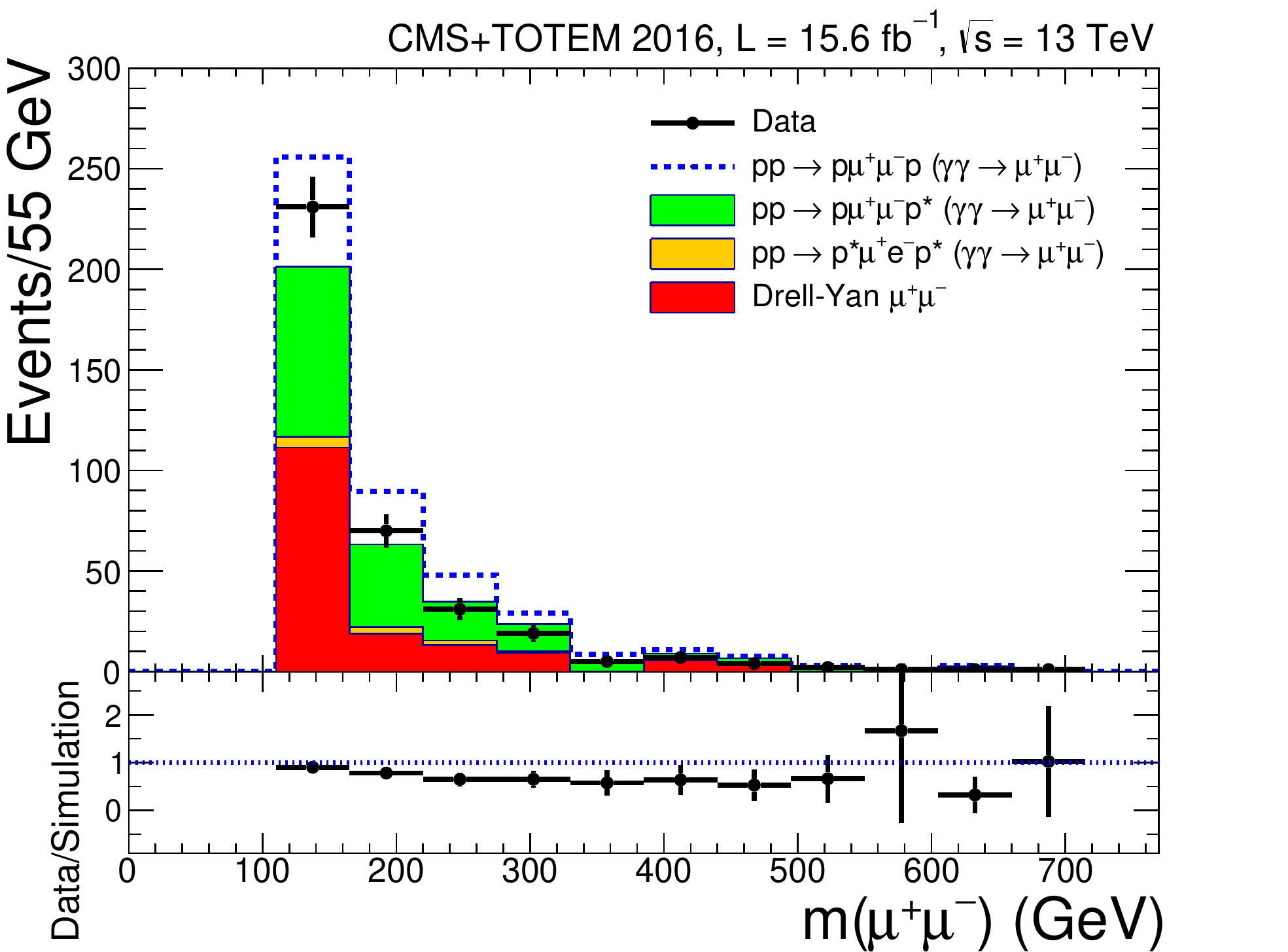}
    \includegraphics[width=.48\textwidth]{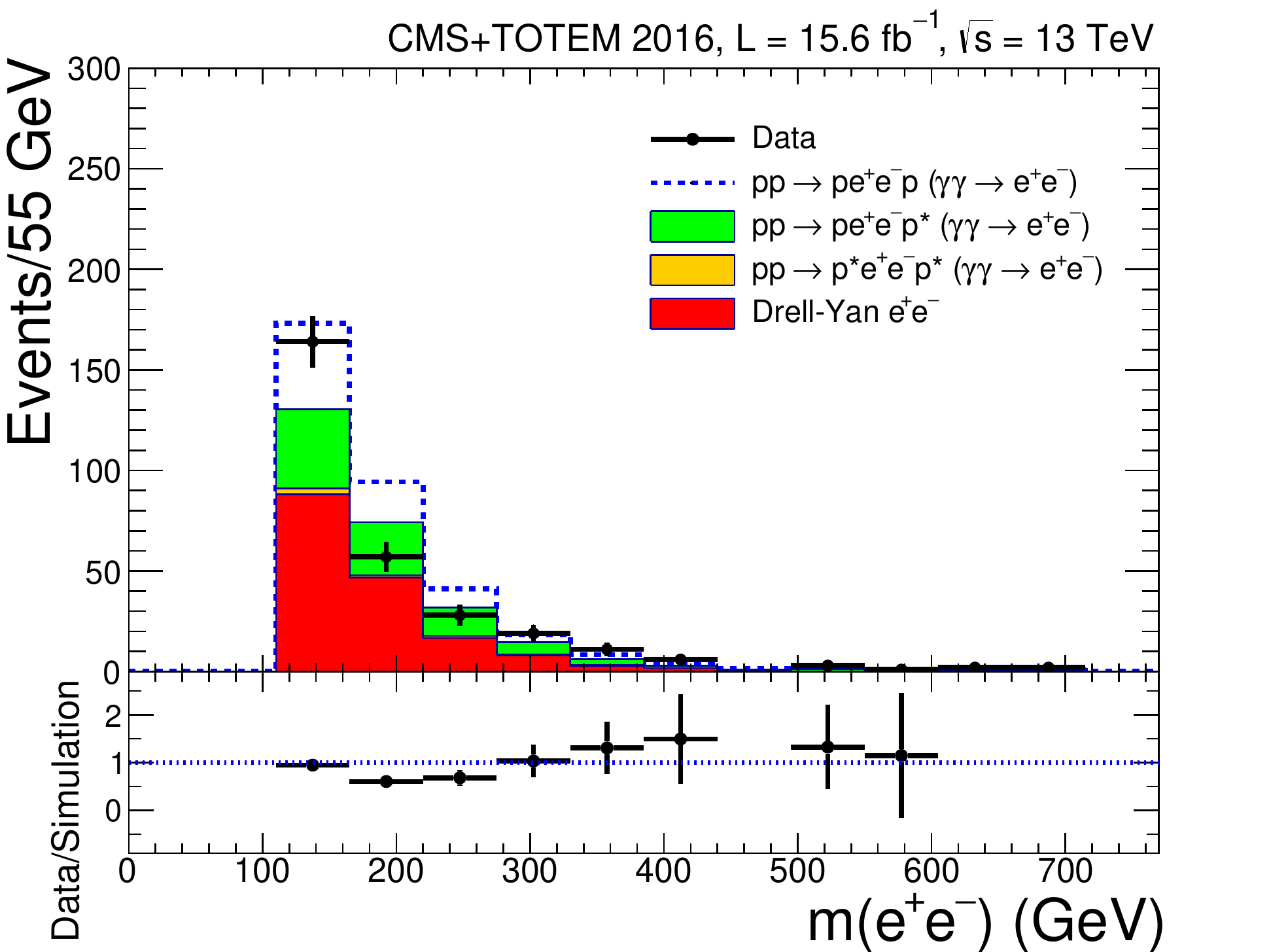}
    \includegraphics[width=.48\textwidth]{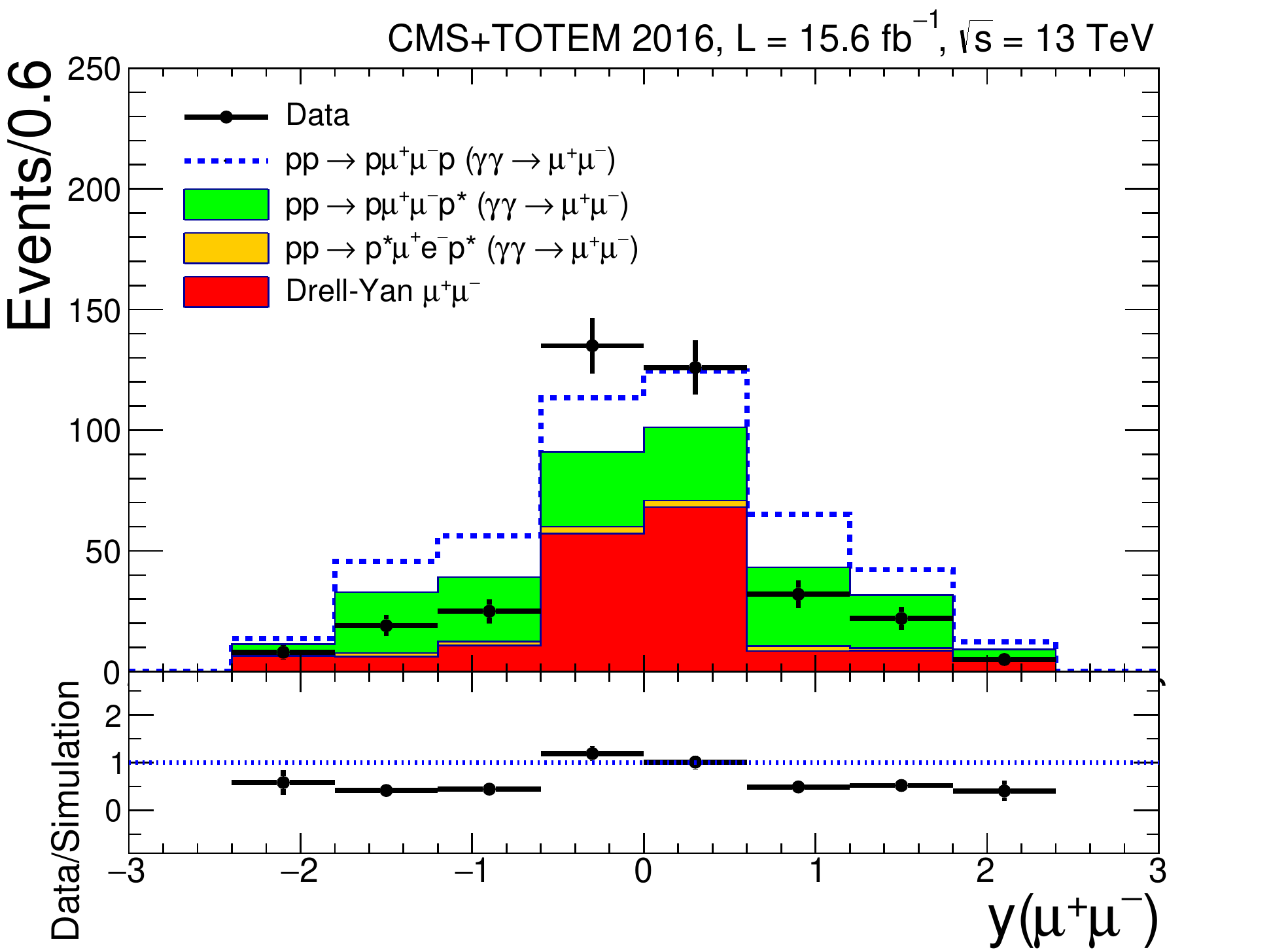}
    \includegraphics[width=.48\textwidth]{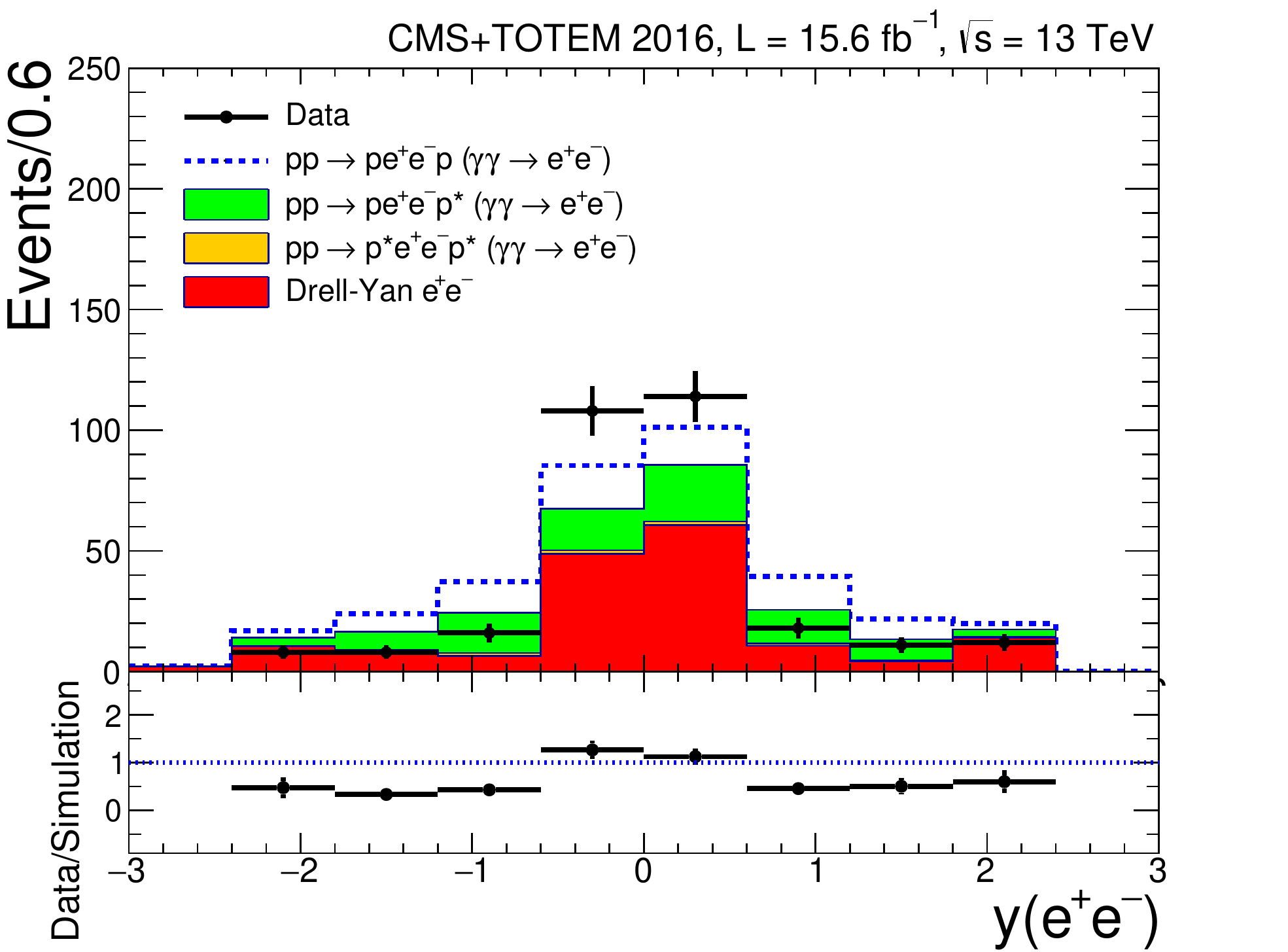}
  \caption{Dimuon (left) and dielectron (right) invariant mass (top) and rapidity (bottom), after all
 central-detector criteria are applied, in $\Pp\Pp$ collisions at 13\TeV. Points with error bars indicate the measured data
(with statistical uncertainties only), and the stacked histograms show the different simulated contributions
for signal and backgrounds (with statistical uncertainty of similar size as
the data). The lower panel in each plot shows the ratio of the data to the sum of all signal and
background predictions.}
  \label{fig:signalregion_plots}
\end{figure}

Figure~\ref{fig:signalregion_plots} shows the distributions of the dimuon
and dielectron invariant mass and rapidity $y$, after all the central
detector requirements just described are applied. The Monte Carlo (MC) predictions are
normalized to the
total integrated luminosity. In addition, for the \textsc{lpair} predictions,
rapidity gap survival probabilities of 0.89, 0.76, and
0.13 are applied to the exclusive, the single
dissociative, and the double dissociative processes, respectively.
The rapidity gap survival probability quantifies the fraction of events in
which no extra soft interactions occur between the colliding protons. These soft interactions produce extra final-state
particles, and thereby suppress the visible (semi)exclusive cross section.
The values used are calculated from modified photon parton distribution functions in the proton that are compatible with
Run 1 LHC measurements. In the case of the proton dissociation processes, these values represent a mix of the incoherent and
QCD evolution terms calculated in Ref.~\cite{Harland-Lang:2016apc}.
This choice of rapidity gap survival probabilities leads to a fair description of
the data for $y$ around zero, but overestimates the results at more forward/backward rapidities, as is clear from
 the bottom panels of Fig.~\ref{fig:signalregion_plots}. A $y$ dependence of the
rapidity gap survival probability is expected in several
models~\cite{Dyndal:2014yea, Harland-Lang:2015cta}.

\subsection{Matching central and proton variables}
\label{central_proton_matching}

Events with at least one well-reconstructed proton track in CT--PPS are retained for further analysis.
For each event, the value of the fractional momentum loss of the scattered proton is estimated from the leptons as:
\begin{equation}
  \xi(\ell^+\ell^-) = \frac{1}{\sqrt{s}} \left[\pt(\ell^{+}) \re^{\pm \eta(\ell^{+})} + \pt(\ell^{-}) \re^{\pm \eta(\ell^{-})}\right],
\end{equation}
where the two solutions for $\pm \eta$ correspond to the protons moving in the ${\pm}z$ direction.

\begin{figure}[h!tb]
  \centering
\includegraphics[width=.9\textwidth]{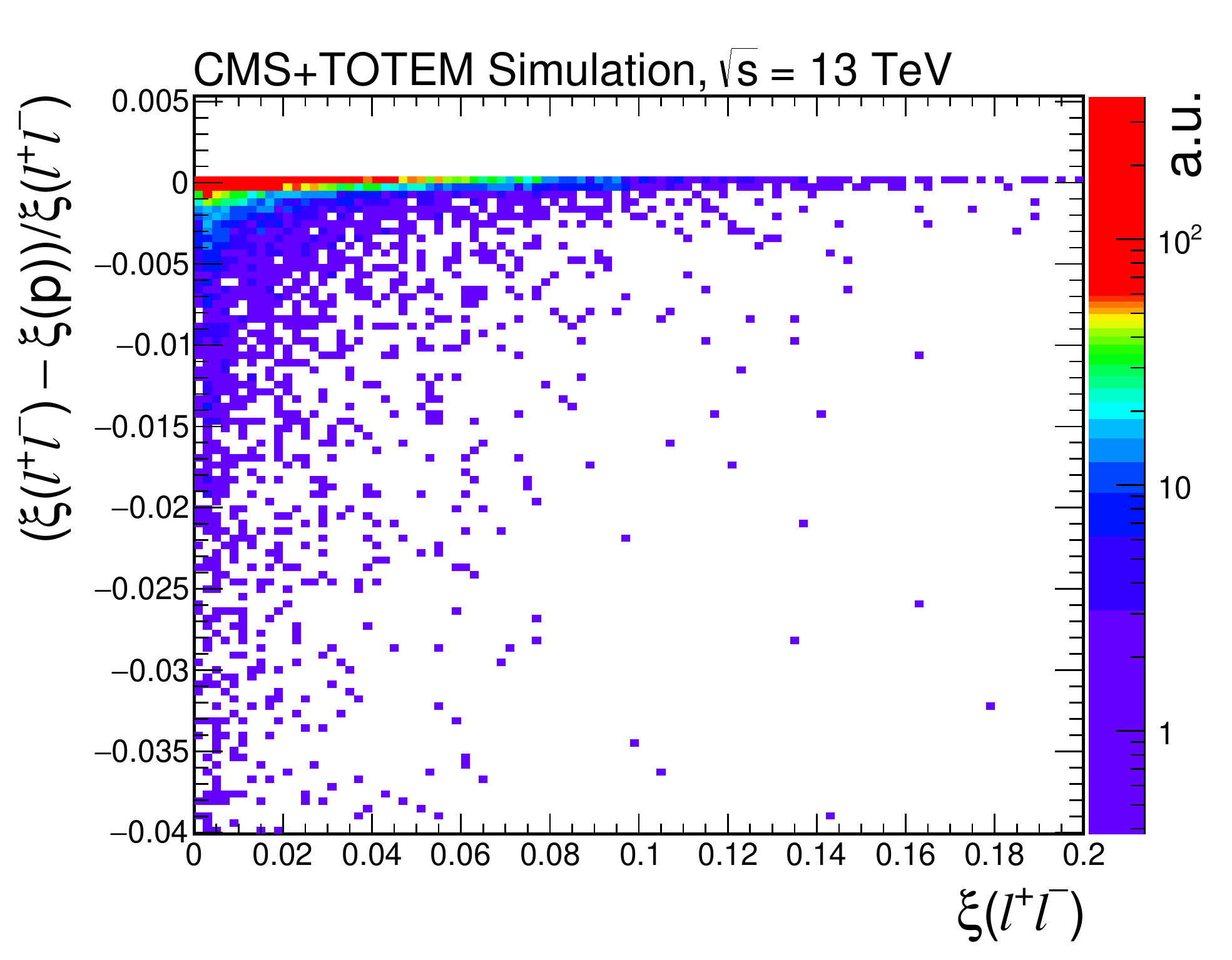}
  \caption{
Generator-level relative difference $(\xi(\ell^+\ell^-) - \xi(\Pp))/(\xi(\ell^+\ell^-))$
vs. $\xi(\ell^+\ell^-)$ for simulated single dissociative
$\PGg\PGg \to \ell^{+}\ell^{-}$ events. Of the two possible
solutions for $\xi(\ell^+\ell^-)$, only the one corresponding to the side with
the intact proton is shown.}
  \label{fig:ximatch_genlevel}
\end{figure}

The formula is exact for exclusive events, but holds also for the single-dissociation case, as illustrated with~\textsc{lpair} simulated events in
Fig.~\ref{fig:ximatch_genlevel}; in this case only one of the two possible
solutions will correspond to the direction of the intact proton. Studies
with~\textsc{lpair} indicate that a mass of the dissociating system larger than about $400\GeV$ is
needed in order to produce a deviation comparable to the expected
$\xi(\ell^{+}\ell^{-})$ resolution of about 3\% (4\%) for dimuons (dielectrons). The latter is obtained from
simulation, with an additional smearing to account for residual
data--simulation differences. The \textsc{lpair} simulation also indicates that the minimum mass of the
dissociating system required to generate activity in the CMS tracker is
about $50\GeV$; the fraction of dissociative events above this threshold is of a few percent.

To be considered as signal candidates, events are further required to have a
value of $\xi(\ell^{+}\ell^{-})$ within the CT--PPS coverage. The minimum
value of $\xi$ observed in an inclusive sample of dilepton-triggered
events, with no selection to enhance $\PGg\PGg$ production, is used.
Numerically this corresponds to:
\begin{itemize}
  \item sector 45, RP 210N: $\xi > $ 0.033,
  \item sector 45, RP 210F:  $\xi > $ 0.024,
  \item sector 56, RP 210N: $\xi > $ 0.042,
  \item sector 56, RP 210F:  $\xi > $ 0.032.
\end{itemize}
The difference between the $\xi$ coverage in the sectors 45 and 56 is due to the asymmetric beam optics.

Finally, the signal region is defined by requiring that $\xi(\ell^+\ell^-)$ and
the corresponding value measured with CT--PPS, $\xi$(RP), agree within
$2\,\sigma$ of the combined uncertainty on $\xi(\ell^+\ell^-)$ and $D_{x}$.

\section{Backgrounds}
\label{backgrounds}

After all selection criteria discussed above, the backgrounds are expected to
arise mainly from prompt $\ell^{+}\ell^{-}$ production combined with proton
tracks from unrelated pileup interactions or beam backgrounds in the same
bunch crossing. The largest background sources of prompt $\ell^{+}\ell^{-}$
production are the Drell--Yan process and $\PGg\PGg \to
\ell^{+}\ell^{-}$ production in which both protons dissociate.

To estimate both the Drell--Yan and the double dissociative backgrounds,
samples of RP tracks from $\Z \to \MM$ and $\Z \to \EE$
events in data are used (referred to as ``$\Z$ control samples" in the following).
For the double dissociative background estimate, \textsc{lpair} simulated events are
also used, in conjunction with the RP tracks from the $\Z$ control samples. To avoid
statistical correlations between the two estimates, only every second event from each sample
is considered for the Drell--Yan estimate, and the remaining part for the double-dissociative
background. In both cases, the background estimation is mostly
based on data, and does not require detailed knowledge of the RP acceptance
and detector efficiency. The procedure is described in the following.

\begin{itemize}

\item  The extra track and acoplanarity selection criteria are not applied to estimate
the Drell--Yan background. Instead, an invariant mass window of $80 < m(\ell^+\ell^-) < 110\GeV$
is imposed, resulting in a high purity sample of Drell--Yan events. A subsample is then selected with
a proton track matching the kinematics of the $\ell^+\ell^-$ pair. The
Drell--Yan events in this subsample tend to be concentrated at
midrapidity, which causes a distortion of the $\xi(\ell^{+}\ell^{-})$
distribution. The distribution is therefore reweighted to match the shape
predicted by the Drell--Yan simulation for events entering the signal
region, with $m(\ell^+\ell^-) > 110\GeV$. Finally, the simulated Drell--Yan sample is
used to obtain the number of matching events expected to pass the track multiplicity, acoplanarity,
and $m(\ell^+\ell^-)$ requirements, given the number observed in the $\Z$ boson control sample.

\item In the case of the background from double dissociation dilepton
production, simulated double dissociation \textsc{lpair} events are randomly mixed
with the background-dominated sample of protons from the $\Z$ boson control sample. The protons from this sample are
used for convenience, and any other sample of protons could have been used; the
information from the central part of the event is not necessary for the present study.

The MC events passing the central detector requirements
are selected, and an exponential function is fitted to the corresponding
$\xi(\ell^+\ell^-)$ distribution. Then a fast simulation is performed in which the
fit is sampled, and the value of $\xi(\ell^+\ell^-)$ is randomly assigned to a
proton from the $\Z$ boson sample.

The background estimate is obtained from the number of events in the fast
simulation that pass the proton selection (cf.
Section~\ref{central_proton_matching}) in addition to the central detector requirements,
normalized to the number of MC events passing the central signal
selection. The procedure just described forces all double dissociation
events to have a background proton in CT--PPS. The background estimate thus
needs to be scaled by the fraction of events passing the central selection
that do not have a proton in CT--PPS. This is obtained from the data.

For the simulation of the double dissociation process, the $y$-independent
rapidity gap survival probability of 0.13 quoted above is used. If instead
the $y$-dependent rapidity gap survival probability discussed in
Section~\ref{EventSelection} were used, the dissociative background and
total background estimates would decrease. The present estimate is thus
conservative.

\item The dissociating system may
contain a final-state proton that falls within the CT--PPS acceptance, even
without overlap of an unrelated proton. However, the simulation indicates
that the total number of such events within the acceptance is negligible.

\end{itemize}

The numbers of background events expected with tracks in either or both of the near and
far RPs in each arm are shown in
Tables~\ref{tab:BackgroundsMuMu}--\ref{tab:BackgroundsEE}.
A total of $11.0 \pm 0.2 \stat$ dimuon events and $10.5 \pm 0.2 \stat$ dielectron
ones are expected within the acceptance, but outside the $2\,\sigma$
matching window. Within the $2\,\sigma$ matching window, the total background prediction is $1.49 \pm
0.07 \stat$ dimuon events and $2.36 \pm 0.09 \stat$ dielectron events
with a matching track in at least one RP, in both arms combined.

\begin{table}
\centering
\topcaption{Estimated backgrounds from Drell--Yan and double-dissociation
\MM production, within the acceptance of at least one of the RPs of a given arm, and in the subsample with
proton kinematics matching within $2\,\sigma$. The bottom row
indicates the total background from the sum of Drell--Yan and double dissociation events.}

\begin{tabular}{lcc}
Arm and background source & Full & $2\,\sigma$\\\hline
Left Drell--Yan & 6.14 $\pm$ 0.13 & 0.75 $\pm$ 0.05 \\
Right Drell--Yan & 5.22 $\pm$ 0.12 & 0.63 $\pm$ 0.04 \\[\cmsTabSkip]
Total Drell--Yan & 11.36 $\pm$ 0.18 & 1.38 $\pm$ 0.06 \\[\cmsTabSkip]
Left double dissociation & 0.57 $\pm$ 0.01 & 0.046 $\pm$ 0.003 \\
Right double dissociation & 0.60 $\pm$ 0.01 & 0.062 $\pm$ 0.004 \\[\cmsTabSkip]
Total double dissociation & 1.17 $\pm$ 0.02 & 0.108 $\pm$ 0.005 \\[\cmsTabSkip]
Total background & 12.52 $\pm$ 0.18 & 1.49 $\pm$ 0.07 \\
\end{tabular}

\label{tab:BackgroundsMuMu}
\end{table}

\begin{table}
\centering
\topcaption{Estimated backgrounds from Drell--Yan and double-dissociation
\EE production, within the acceptance of at least one of the RPs of a given arm, and in the subsample with
proton kinematics matching within $2\,\sigma$. The bottom row
indicates the total background from the sum of Drell--Yan and double dissociation events.}

\begin{tabular}{lcc}
Arm and background source & Full & $2\,\sigma$ \\\hline
Left Drell--Yan & 6.24 $\pm$ 0.13 & 1.07 $\pm$ 0.06 \\
Right Drell--Yan & 6.09 $\pm$ 0.14 & 1.23 $\pm$ 0.06 \\[\cmsTabSkip]
Total Drell--Yan & 12.33 $\pm$ 0.19 & 2.30 $\pm$ 0.09 \\[\cmsTabSkip]
Left double dissociation & 0.31 $\pm$ 0.01 & 0.035 $\pm$ 0.002 \\
Right double dissociation & 0.25 $\pm$ 0.01 & 0.032 $\pm$ 0.002 \\[\cmsTabSkip]
Total double dissociation & 0.56 $\pm$ 0.01 & 0.067 $\pm$ 0.003 \\[\cmsTabSkip]
Total background & 12.89 $\pm$ 0.19 & 2.36 $\pm$ 0.09 \\
\end{tabular}

\label{tab:BackgroundsEE}
\end{table}

\begin{table}[]
\centering
\topcaption{Sources of systematic uncertainties in the estimates of Drell--Yan and double-dissociation backgrounds in the dimuon and dielectron channels.}
\label{tab:syst_table}
\begin{tabular}{l|cccc}

                                                & \multicolumn{2}{c}{$\MM$}                & \multicolumn{2}{c}{$\EE$}                    \\\hline
\multicolumn{1}{c|}{Sources of uncertainty}    & \multicolumn{1}{c}{Drell--Yan} & Double diss.        & \multicolumn{1}{c}{Drell--Yan} & Double diss.        \\[\cmsTabSkip]
Statistics of \Z sample                          & 5\%                            & 5\%                 & 4\%                            & 4\%                 \\
$\xi(\ell^{+}\ell^{-})$ reweighting             & 25\%                           & \NA                   & 11\%                           & \NA                   \\
Track multiplicity modeling                    & 28\%                           & \NA                   & 14\%                           & \NA                   \\
Survival probability                            & \NA                              & 100\%               & \NA                              & 100\%               \\
Luminosity                                      & \NA                              & 2.5\%               & \NA                              & 2.5\%               \\
\end{tabular}
\end{table}

The systematic uncertainties in the Drell--Yan and double dissociation backgrounds are shown in Table~\ref{tab:syst_table} and are estimated as follows.
A 5\% contribution is assigned to reflect the statistical uncertainty of the control sample of protons from the $\Z$ boson mass region for the dimuon case, and a 4\% contribution for the dielectron channel.
In addition, the Drell--Yan background estimate is affected by uncertainties related to the reweighting of the $\xi(\ell^{+}\ell^{-})$ distribution and the modeling of the track multiplicity distribution in the simulation.
The former is obtained as the difference of the background estimates with
and without reweighting, leading to a 25\%\,(11\%) relative uncertainty
in the dimuon (dielectron) channel.
The latter is estimated from the difference between data and simulation in the low-multiplicity region, with 1--5 additional tracks near the dilepton vertex, resulting in 28 and 14\% relative uncertainties for the dimuon and dielectron channels, respectively.
The double-dissociation process has never been measured directly, and therefore the background estimate for this process also includes a 100\% relative uncertainty on the rapidity gap survival probability.
Finally, the double-dissociation background includes a 2.5\% integrated luminosity uncertainty~\cite{Lumi} applied to the normalization of the simulated samples.

As a further check of the pileup background estimate, a set of pseudo-experiments is performed in which the measured values of $\xi(\ell^+\ell^-)$ within the CT--PPS acceptance are randomly coupled with $\xi(\Pp)$ values from events without any offline selection imposed on the central variables.
The dilepton system and the proton originate from different events and are thus uncorrelated.
Such a procedure is repeated $10^4$ times, and the average number of events in which $\xi(\ell^+\ell^-)$ and $\xi(\Pp)$ match within $2\,\sigma$ is determined.
The result is consistent with the background estimates of $1.49 \pm 0.07 \stat$ and $2.36 \pm 0.09 \stat$ events discussed above for the dimuon and dielectron channels, respectively.

\section{Results}
\label{results}

In the \MM channel, a total of 17 events are observed with $\xi(\MM)$ within the CT--PPS acceptance, and at least one track detected in the relevant RPs.
Five of those events have a mismatch of ${\geq}2\,\sigma$ between the dimuon and the proton kinematics, compared to $11.0 \pm 4.0$\,(stat+syst) such events expected from background; twelve events have a track in at least one of the two RPs matching $\xi(\MM)$ within $2\,\sigma$.
The significance of observing 12 events over the background estimate of $1.49 \pm 0.07 \stat \pm 0.53 \syst$ is $4.3\,\sigma$, estimated by performing pseudo-experiments according to a Poisson distribution, including the systematic uncertainties profiled as log-normal nuisance parameters.

The invariant masses and rapidities of the \MM candidate events are consistent with the expected single-arm acceptance, given
the LHC optics and the position of the RPs.
No events are observed with matching protons in both arms; the highest-mass event is at $m(\MM) = 342\GeV$, approximately $20\GeV$ below the threshold required to detect both protons.

In the \EE channel, a total of 23 events are observed with $\xi(\EE)$ within the CT--PPS acceptance, and at least one track detected in the relevant RPs.
Fifteen of those events have a mismatch of ${\geq}2\,\sigma$ between $\xi$(RP) and $\xi(\EE)$ compared to the expectation of $10.5 \pm 2.1$ (stat+syst).
Eight events have a scattered proton candidate in at least one of the two RPs matching $\xi(\EE)$ within $2\,\sigma$.
The significance of observing 8 events with a background estimate of $2.36 \pm 0.09 \stat \pm 0.47 \syst$ is $2.6\,\sigma$, including the systematic uncertainties profiled as log-normal nuisance parameters.

As for the dimuon case, no events in the \EE channel are observed with matching protons in both arms, although the highest mass events are at $m(\EE) = 650$ and $917\GeV$, in the region where the double-arm acceptance is nonzero.
Studies based on \textsc{lpair} indicate that less than one exclusive event is expected in this region for the integrated luminosity of the present data.
The data show no activity compatible with a track in the RPs on the side where no proton is observed, thus ruling out track reconstruction problems.
The expected $\xi$ of the missing proton derived from the lepton kinematics corresponds to the detector region well outside the area suffering from inefficiencies induced by radiation damage.
The two events observed are thus likely to be semiexclusive events, or background events with an uncorrelated proton.

Central semiexclusive dilepton events are expected to have very small values of $\abs{t}$, the absolute value of the four-momentum squared exchanged at the proton vertices.
As mentioned earlier, only the vertical component  of the scattering angle, and hence of the proton transverse momentum, is currently measured.
For 11 candidate dimuon events out of the 12, the vertical component of the scattering angle is compatible with zero within at most $2.5\,\sigma$, where $\sigma$ is the uncertainty of the vertical component of the scattering angle.
For one event, the discrepancy is $3.5\,\sigma$, in agreement with the background estimate.
Also for the dielectron data the vertical component of the scattering angle, and hence of the proton transverse momentum, is measured; it is consistent with zero, as expected for the signal, for six of the eight events.
Two events have values more than $3\,\sigma$ away from zero.
This is again consistent with the background estimate. The vertical component of the scattering angle for the two highest-mass $\EE$ events is compatible with zero.

\begin{figure}[h!tb]
  \centering
\includegraphics[width=.45\textwidth]{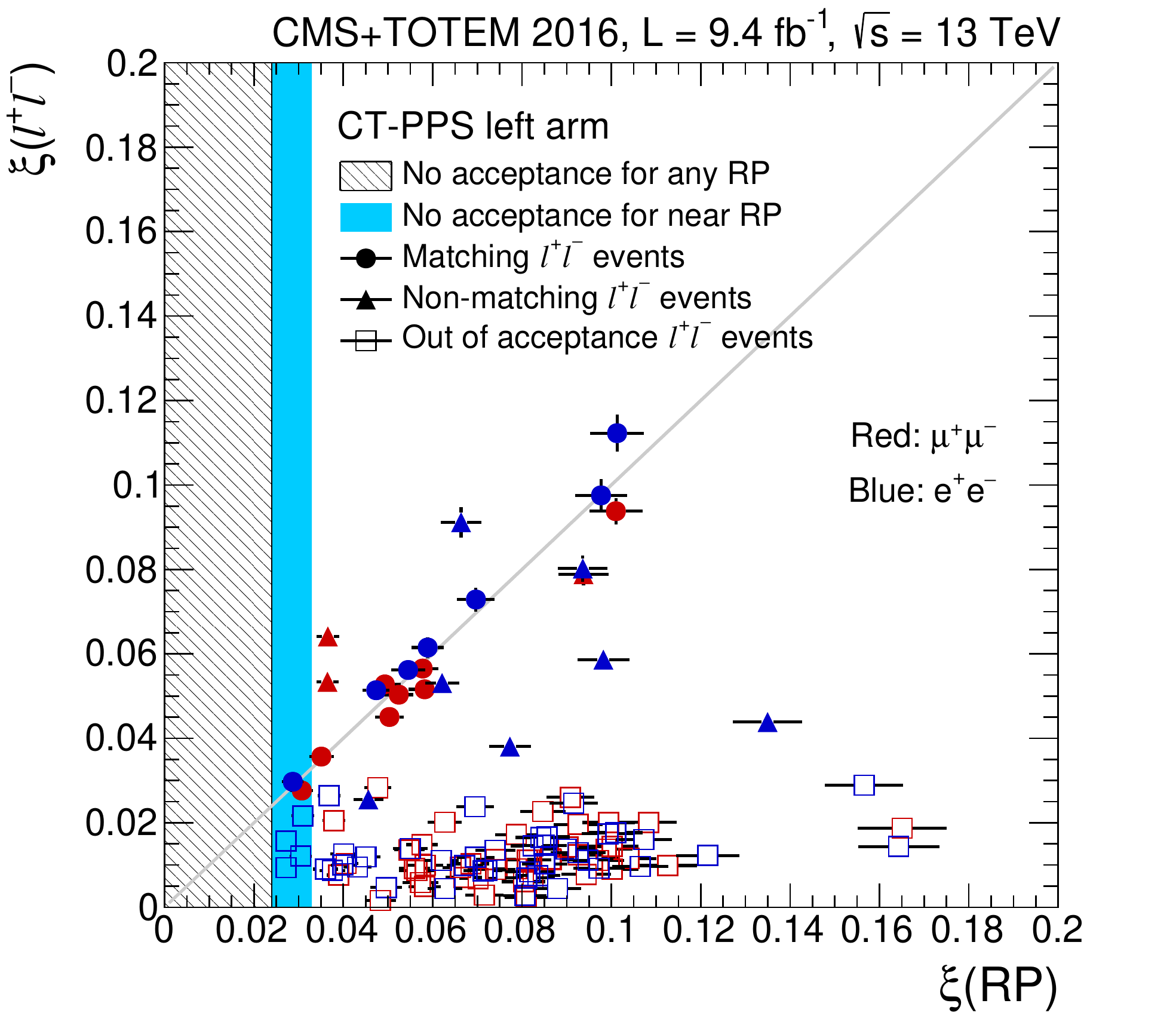}
\includegraphics[width=.45\textwidth]{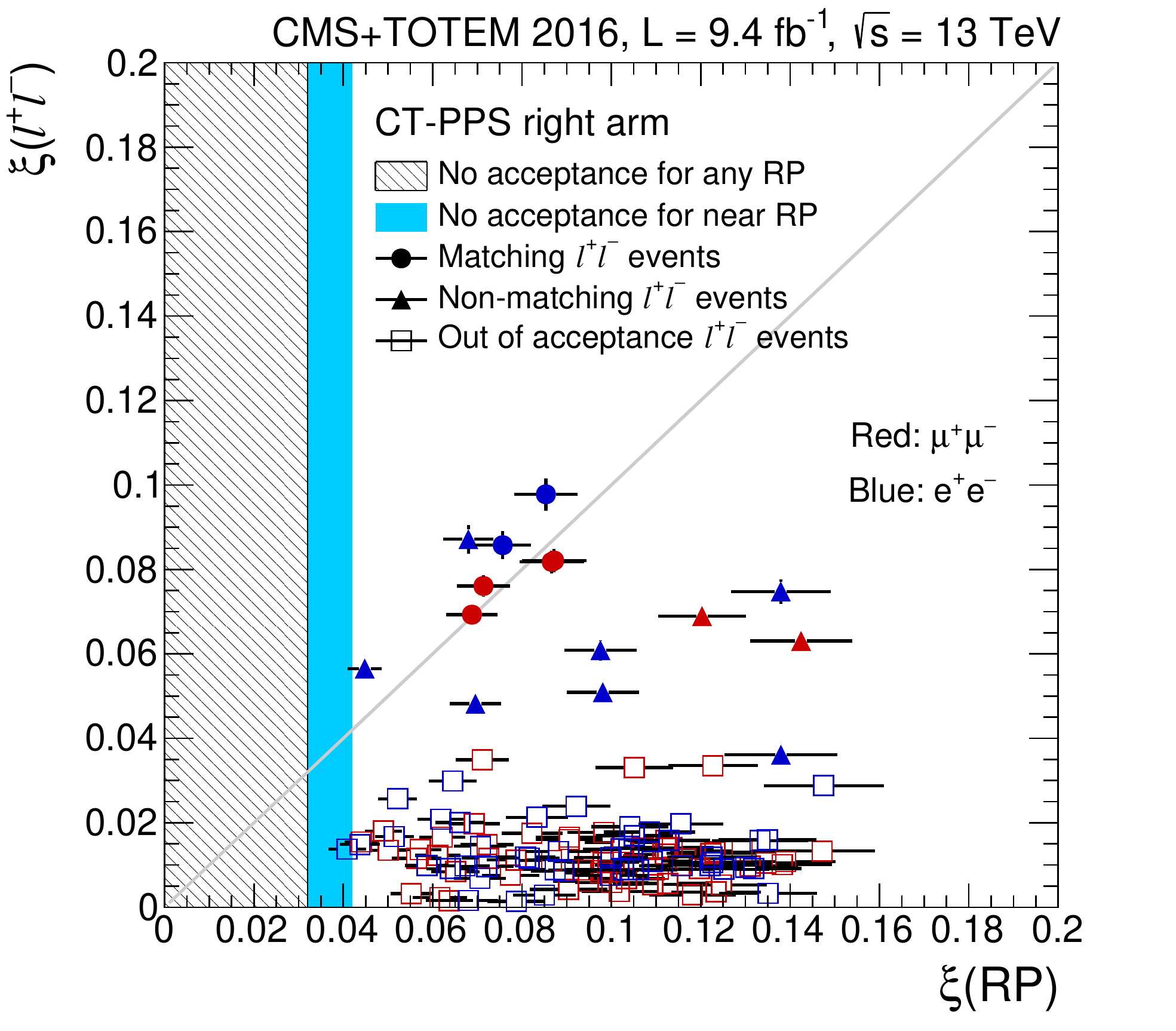}
  \caption{Correlation between the fractional values of the proton momentum loss measured in the central dilepton system, $\xi(\ell^+\ell^-)$, and in the RPs, $\xi$(RP), for
both RPs in each arm combined. The 45 (left) and 56 (right) arms are shown. The hatched region corresponds to the kinematical region outside the acceptance of both the near and far RPs, while the shaded (pale blue) region corresponds to the region outside the acceptance of the near RP.
	For the events in which a track is detected in both, the $\xi$ value measured at the near RP is plotted.
	The horizontal error bars indicate the uncertainty of $\xi$(RP), and the vertical bars the uncertainty of $\xi(\ell^+\ell^-)$.
	The events labeled ``out of acceptance'' are those in which $\xi(\ell^+\ell^-)$ corresponds to a signal proton outside the RP acceptance; in these events a background proton is detected with nonmatching kinematics.}
  \label{fig:corr_plot_combined_mumu_ee}
\end{figure}

\begin{figure}[h!]
  \centering
    \includegraphics[width=.5\textwidth]{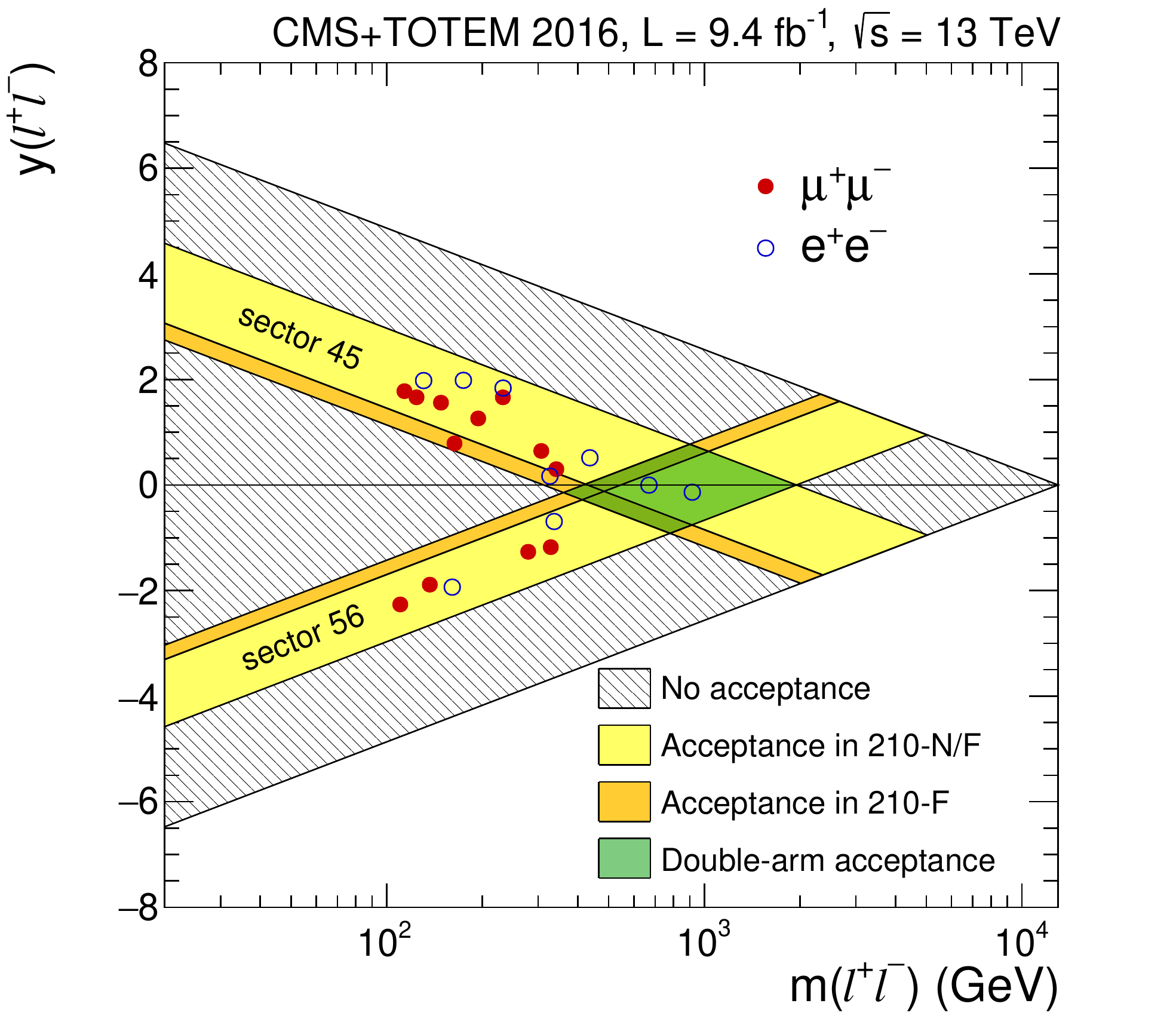}
  \caption{Expected acceptance regions in the rapidity vs. invariant mass plane
overlaid with the observed dimuon (closed circles) and dielectron (open circles) signal candidate
events. The ``double-arm acceptance" refers to exclusive events, $\Pp\Pp \to \Pp \ell^+\ell^- \Pp$.
Following the CMS convention, the positive (negative) rapidity region corresponds
to the 45 (56) LHC sector.}
  \label{fig:MarioPlotData_mumu_ee}
\end{figure}

The correlation of $\xi(\ell^{+}\ell^{-})$ versus $\xi$(RP) and the mass versus rapidity
distributions, for the combined dimuon and dielectron results, are shown in Figs.~\ref{fig:corr_plot_combined_mumu_ee}
and~\ref{fig:MarioPlotData_mumu_ee}. The combined signal significance is estimated by performing pseudo-experiments according to a
joint distribution, including systematic uncertainties, and corresponds to an excess of $5.1\,\sigma$ over the background. In the calculation, the
uncertainty on the integrated luminosity and that on the
rapidity gap survival probability are assumed to be fully correlated between the
two channels. All other sources are taken as independent.
Of the 20 total events selected, 13 have a track in both the near and far RPs. In
these events, the two independent $\xi$ measurements agree within 4\%.

The fractions of the exclusive and single proton dissociative contributions in the final sample of matching events are estimated by comparing their acoplanarity distribution to those expected for the two classes of events in \textsc{lpair}.
This results in a contribution of approximately 70\% from single proton dissociation, consistent within large uncertainties with the predictions of \textsc{lpair} weighted by the rapidity gap survival probabilities.
The dominance of single dissociation is also consistent with the lack of a second observed proton in the two high-mass \EE events.

The observed yields are consistent with those predicted by \textsc{lpair}
modified by the rapidity gap survival probabilities, assuming the fraction of
single proton dissociation events from the acoplanarity comparison just
discussed. The full simulation of the CMS central apparatus (Section~\ref{datasets}) is used. For
the scattered protons, the prediction includes the effect of the CT--PPS acceptance,
that of radiation damage in the silicon strip sensors, and the inefficiency due
to multiple proton tracks. The comparison is performed in the region where
radiation damage is less severe, corresponding to $\xi\text{(RP)}\geq 0.05$.

\section{Summary}
\label{summary}

We have studied $\PGg\PGg \to \MM$ and
$\PGg\PGg \to \EE$ production together with forward protons
reconstructed in the CMS--TOTEM precision proton spectrometer (CT--PPS),
using a sample of 9.4\fbinv collected in proton--proton collisions at $\sqrt{s} = 13\TeV$. The Roman Pot alignment and LHC optics
corrections have been determined using a high statistics sample of forward
protons. A total of 12 $\PGg\PGg \to \MM$ and 8
$\PGg\PGg \to \EE$ events are observed with
dilepton invariant mass larger than 110\GeV, and a forward proton with consistent kinematics. This corresponds to an excess larger than five standard
deviations over the expected background from double-dissociative and Drell--Yan dilepton processes. The result represents the first observation
of proton-tagged $\PGg\PGg$ collisions at the electroweak scale.
The present data demonstrate the excellent performance of CT--PPS and its
potential for high-mass exclusive (proton-tagged) measurements. With its 2016 operation, CT--PPS has proven for the first time
the feasibility of continuously operating a near-beam proton spectrometer
at a high-luminosity hadron collider.

\begin{acknowledgments}
\hyphenation{Bundes-ministerium Forschungs-gemeinschaft Forschungs-zentren Rachada-pisek} We congratulate our colleagues in the CERN accelerator departments for the excellent performance of the LHC and thank the technical and administrative staffs at CERN and at other CMS and TOTEM institutes for their contributions to the success of the CMS and TOTEM effort. In addition, we gratefully acknowledge the computing centers and personnel of the Worldwide LHC Computing Grid for delivering so effectively the computing infrastructure essential to our analyses. Finally, we acknowledge the enduring support for the construction and operation of the LHC and the CMS and TOTEM detectors provided by the following funding agencies: the Austrian Federal Ministry of Science, Research and Economy and the Austrian Science Fund; the Belgian Fonds de la Recherche Scientifique, and Fonds voor Wetenschappelijk Onderzoek; the Brazilian Funding Agencies (CNPq, CAPES, FAPERJ, and FAPESP); the Bulgarian Ministry of Education and Science; CERN; the Chinese Academy of Sciences, Ministry of Science and Technology, and National Natural Science Foundation of China; the Colombian Funding Agency (COLCIENCIAS); the Croatian Ministry of Science, Education and Sport, and the Croatian Science Foundation; the Research Promotion Foundation, Cyprus; the Secretariat for Higher Education, Science, Technology and Innovation, Ecuador; the Ministry of Education and Research, Estonian Research Council via IUT23-4 and IUT23-6 and European Regional Development Fund, Estonia; the Academy of Finland, Finnish Ministry of Education and Culture, Helsinki Institute of Physics,  the Magnus Ehrnrooth Foundation, the Waldemar von Frenckell Foundation, and the Finnish Academy of Science and Letters (The Vilho Yrj{\"o} and Kalle V{\"a}is{\"a}l{\"a} Fund); the Institut National de Physique Nucl\'eaire et de Physique des Particules~/~CNRS, and Commissariat \`a l'\'Energie Atomique et aux \'Energies Alternatives~/~CEA, France; the Bundesministerium f\"ur Bildung und Forschung, Deutsche Forschungsgemeinschaft, and Helmholtz-Gemeinschaft Deutscher Forschungszentren, Germany; the General Secretariat for Research and Technology, Greece; the National Scientific Research Foundation, and National Innovation Office, the OTKA NK 101438, and the EFOP-3.6.1-16-2016-00001 grant (Hungary); the Department of Atomic Energy and the Department of Science and Technology, India; the Institute for Studies in Theoretical Physics and Mathematics, Iran; the Science Foundation, Ireland; the Istituto Nazionale di Fisica Nucleare, Italy; the Ministry of Science, ICT and Future Planning, and National Research Foundation (NRF), Republic of Korea; the Lithuanian Academy of Sciences; the Ministry of Education, and University of Malaya (Malaysia); the Mexican Funding Agencies (BUAP, CINVESTAV, CONACYT, LNS, SEP, and UASLP-FAI); the Ministry of Business, Innovation and Employment, New Zealand; the Pakistan Atomic Energy Commission; the Ministry of Science and Higher Education and the National Science Center, Poland; the Funda\c{c}\~ao para a Ci\^encia e a Tecnologia, Portugal; JINR, Dubna; the Ministry of Education and Science of the Russian Federation, the Federal Agency of Atomic Energy of the Russian Federation, Russian Academy of Sciences, the Russian Foundation for Basic Research and the Russian Competitiveness Program of NRNU ``MEPhI''; the Ministry of Education, Science and Technological Development of Serbia; the Secretar\'{\i}a de Estado de Investigaci\'on, Desarrollo e Innovaci\'on, Programa Consolider-Ingenio 2010, Plan de Ciencia, Tecnolog\'{i}a e Innovaci\'on 2013-2017 del Principado de Asturias and Fondo Europeo de Desarrollo Regional, Spain; the Swiss Funding Agencies (ETH Board, ETH Zurich, PSI, SNF, UniZH, Canton Zurich, and SER); the Ministry of Science and Technology, Taipei; the Thailand Center of Excellence in Physics, the Institute for the Promotion of Teaching Science and Technology of Thailand, Special Task Force for Activating Research and the National Science and Technology Development Agency of Thailand; the Scientific and Technical Research Council of Turkey, and Turkish Atomic Energy Authority; the National Academy of Sciences of Ukraine, and State Fund for Fundamental Researches, Ukraine; the Science and Technology Facilities Council, UK; the US Department of Energy, and the US National Science Foundation.

Individuals have received support from the Marie-Curie program and the European Research Council and Horizon 2020 Grant, contract No. 675440 (European Union); the Leventis Foundation; the A. P. Sloan Foundation; the Alexander von Humboldt Foundation; the Belgian Federal Science Policy Office; the Fonds pour la Formation \`a la Recherche dans l'Industrie et dans l'Agriculture (FRIA-Belgium); the Agentschap voor Innovatie door Wetenschap en Technologie (IWT-Belgium); the F.R.S.-FNRS and FWO (Belgium) under the ``Excellence of Science - EOS'' - be.h project n. 30820817; the Ministry of Education, Youth and Sports (MEYS) and MSMT CR of the Czech Republic; Nylands nation vid Helsingfors Universitet (Finland);
the J\'{a}nos Bolyai Research Scholarship of the Hungarian Academy of Sciences; the NKP-17-4 New National Excellence Program of the Hungarian Ministry of Human Capacities; the Council of Scientific and Industrial Research, India; the HOMING PLUS program of the Foundation for Polish Science, cofinanced from European Union, Regional Development Fund, the Mobility Plus program of the Ministry of Science and Higher Education, the National Science Center (Poland), contracts Harmonia 2014/14/M/ST2/00428, Opus 2014/13/B/ST2/02543, 2014/15/B/ST2/03998, and 2015/19/B/ST2/02861, Sonata-bis 2012/07/E/ST2/01406; the National Priorities Research Program by Qatar National Research Fund; the Programa Severo Ochoa del Principado de Asturias; the Thalis and Aristeia programs cofinanced by EU-ESF and the Greek NSRF; the Rachadapisek Sompot Fund for Postdoctoral Fellowship, Chulalongkorn University and the Chulalongkorn Academic into Its 2nd Century Project Advancement Project (Thailand); the Welch Foundation, contract C-1845; and the Weston Havens Foundation (USA).
\end{acknowledgments}

\bibliography{auto_generated}

\cleardoublepage \appendix\section{The CMS Collaboration \label{app:collab}}\begin{sloppypar}\hyphenpenalty=5000\widowpenalty=500\clubpenalty=5000\input{PPS-17-001-authorlist.tex}\section{The TOTEM Collaboration\label{app:totem}}\input{authorlist_data_updated_CTPPS}\end{sloppypar}
\end{document}

%% file: PPS-17-001-authorlist.tex
\vskip\cmsinstskip
\textbf{Yerevan~Physics~Institute, Yerevan, Armenia}\\*[0pt]
A.M.~Sirunyan, A.~Tumasyan
\vskip\cmsinstskip
\textbf{Institut~f\"{u}r~Hochenergiephysik, Wien, Austria}\\*[0pt]
W.~Adam, F.~Ambrogi, E.~Asilar, T.~Bergauer, J.~Brandstetter, E.~Brondolin, M.~Dragicevic, J.~Er\"{o}, A.~Escalante~Del~Valle, M.~Flechl, M.~Friedl, R.~Fr\"{u}hwirth\cmsAuthorMark{1}, V.M.~Ghete, J.~Grossmann, J.~Hrubec, M.~Jeitler\cmsAuthorMark{1}, A.~K\"{o}nig, N.~Krammer, I.~Kr\"{a}tschmer, D.~Liko, T.~Madlener, I.~Mikulec, E.~Pree, N.~Rad, H.~Rohringer, J.~Schieck\cmsAuthorMark{1}, R.~Sch\"{o}fbeck, M.~Spanring, D.~Spitzbart, A.~Taurok, W.~Waltenberger, J.~Wittmann, C.-E.~Wulz\cmsAuthorMark{1}, M.~Zarucki
% ...
\vskip\cmsinstskip
% ...
1:~Also at~Vienna~University~of~Technology, Vienna, Austria\\
% ...

%% file: authorlist_data_updated_CTPPS.tex
\newcommand{\AddAuthor}[2]{#1$^{#2}$,\ }
\newcommand\AddAuthorLast[2]{#1$^{#2}$}
\noindent
    \AddAuthor{G.~Antchev}{a}%
	\AddAuthor{P.~Aspell}{9}%
	\AddAuthor{I.~Atanassov}{a}%
	\AddAuthor{V.~Avati}{7,9}%
	\AddAuthor{J.~Baechler}{9}%
	\AddAuthor{C.~Baldenegro~Barrera}{11}%
	\AddAuthor{V.~Berardi}{4a,4b}%
	\AddAuthor{M.~Berretti}{2a}%
	\AddAuthor{E.~Bossini}{9,6b}%
	\AddAuthor{U.~Bottigli}{6b}%
	\AddAuthor{M.~Bozzo}{5a,5b}%
	\AddAuthor{F.~S.~Cafagna}{4a}%
	\AddAuthor{M.~G.~Catanesi}{4a}%
	\AddAuthor{M.~Csan\'{a}d}{3a,b}%
	\AddAuthor{T.~Cs\"{o}rg\H{o}}{3a,3b}%
	\AddAuthor{M.~Deile}{9}%
	\AddAuthor{F.~De~Leonardis}{4c,4a}%
	\AddAuthor{A.~D'Orazio}{4c,4a}%
	\AddAuthor{M.~Doubek}{1c}%
	\AddAuthor{D.~Druzhkin}{9}%
	\AddAuthor{K.~Eggert}{10}%
	\AddAuthor{V.~Eremin}{e}%
	\AddAuthor{A.~Fiergolski}{9}%
	\AddAuthor{F.~Garcia}{2a}%
	\AddAuthor{V.~Georgiev}{1a}%
	\AddAuthor{S.~Giani}{9}%
	\AddAuthor{L.~Grzanka}{7,c}%
	\AddAuthor{J.~Hammerbauer}{1a}%
	\AddAuthor{J.~Heino}{2a}%
	\AddAuthor{T.~Isidori}{11}%
	\AddAuthor{V.~Ivanchenko}{8}%
	\AddAuthor{M.~Janda}{1c}%
	\AddAuthor{A.~Karev}{9}%
	\AddAuthor{J.~Ka\v{s}par}{6a,1b}%
	\AddAuthor{J.~Kopal}{9}%
	\AddAuthor{V.~Kundr\'{a}t}{1b}%
	\AddAuthor{S.~Lami}{6a}%
	\AddAuthor{G.~Latino}{6b}%
	\AddAuthor{R.~Lauhakangas}{2a}%
	\AddAuthor{R.~Linhart}{1a}%
	\AddAuthor{C.~Lindsey}{11}%
	\AddAuthor{M.~V.~Lokaj\'{\i}\v{c}ek}{1b}%
	\AddAuthor{L.~Losurdo}{6b}%
	\AddAuthor{M.~Lo~Vetere}{5b,5a,\dagger}%
	\AddAuthor{F.~Lucas~Rodr\'{i}guez}{9}%
	\AddAuthor{M.~Macr\'{\i}}{5a}%
	\AddAuthor{M.~Malawski}{7,c}%
	\AddAuthor{N.~Minafra}{11}%
	\AddAuthor{S.~Minutoli}{5a}%
	\AddAuthor{T.~Naaranoja}{2a,2b}%
	\AddAuthor{F.~Nemes}{9,3a}%
	\AddAuthor{H.~Niewiadomski}{10}%
	\AddAuthor{T.~Nov\'{a}k}{3b}%
	\AddAuthor{E.~Oliveri}{9}%
	\AddAuthor{F.~Oljemark}{2a,2b}%
	\AddAuthor{M.~Oriunno}{f}%
	\AddAuthor{K.~\"{O}sterberg}{2a,2b}%
	\AddAuthor{P.~Palazzi}{9}%
	\AddAuthor{V.~Passaro}{4c,4a}%
	\AddAuthor{Z.~Peroutka}{1a}%
	\AddAuthor{J.~Proch\'{a}zka}{1b}%
	\AddAuthor{M.~Quinto}{4a,4b}%
	\AddAuthor{E.~Radermacher}{9}%
	\AddAuthor{E.~Radicioni}{4a}%
	\AddAuthor{F.~Ravotti}{9}%
	\AddAuthor{G.~Ruggiero}{9}%
	\AddAuthor{H.~Saarikko}{2a,2b}%
	\AddAuthor{A.~Scribano}{6a}%
	\AddAuthor{J.~Siroky}{1a}%
	\AddAuthor{J.~Smajek}{9}%
	\AddAuthor{W.~Snoeys}{9}%
	\AddAuthor{R.~Stefanovitch}{9}%
	\AddAuthor{J.~Sziklai}{3a}%
	\AddAuthor{C.~Taylor}{10}%
	\AddAuthor{E.~Tcherniaev}{8}%
	\AddAuthor{N.~Turini}{6b}%
	\AddAuthor{V.~Vacek}{1c}%
	\AddAuthor{J.~Welti}{2a,2b}%
	\AddAuthor{J.~Williams}{11}%
	\AddAuthor{P.~Wyszkowski}{7}%
	\AddAuthor{J.~Zich}{1a}%
	\AddAuthorLast{K.~Zielinski}{7}%

\vskip 4pt plus 4pt
\let\thefootnote\relax
\newcommand{\AddInstitute}[2]{${}^{#1}$#2\\}
\newcommand{\AddExternalInstitute}[2]{\footnote{${}^{#1}$ #2}}
\noindent
	\AddInstitute{1a}{University of West Bohemia, Pilsen, Czech Republic.}
	\AddInstitute{1b}{Institute of Physics of the Academy of Sciences of the Czech Republic, Prague, Czech Republic.}
	\AddInstitute{1c}{Czech Technical University, Prague, Czech Republic.}
	\AddInstitute{2a}{Helsinki Institute of Physics, University of Helsinki, Helsinki, Finland.}
	\AddInstitute{2b}{Department of Physics, University of Helsinki, Helsinki, Finland.}
	\AddInstitute{3a}{Wigner Research Centre for Physics, RMKI, Budapest, Hungary.}
	\AddInstitute{3b}{EKU KRC, Gy\"ongy\"os, Hungary.}
	\AddInstitute{4a}{INFN Sezione di Bari, Bari, Italy.}
	\AddInstitute{4b}{Dipartimento Interateneo di Fisica di Bari, Bari, Italy.}
	\AddInstitute{4c}{Dipartimento di Ingegneria Elettrica e dell'Informazione --- Politecnico di Bari, Bari, Italy.}
	\AddInstitute{5a}{INFN Sezione di Genova, Genova, Italy.}
	\AddInstitute{5b}{Universit\`{a} degli Studi di Genova, Italy.}
	\AddInstitute{6a}{INFN Sezione di Pisa, Pisa, Italy.}
	\AddInstitute{6b}{Universit\`{a} degli Studi di Siena and Gruppo Collegato INFN di Siena, Siena, Italy.}
	\AddInstitute{7}{AGH University of Science and Technology, Krakow, Poland.}
	\AddInstitute{8}{Tomsk State University, Tomsk, Russia.}
	\AddInstitute{9}{CERN, Geneva, Switzerland.}
	\AddInstitute{10}{Case Western Reserve University, Dept.~of Physics, Cleveland, OH, USA.}
	\AddInstitute{11}{The University of Kansas, Lawrence, USA.}


	\AddExternalInstitute{a}{INRNE-BAS, Institute for Nuclear Research and Nuclear Energy, Bulgarian Academy of Sciences, Sofia, Bulgaria.}
	\AddExternalInstitute{b}{Department of Atomic Physics, ELTE University, Budapest, Hungary.}
	\AddExternalInstitute{c}{Institute of Nuclear Physics, Polish Academy of Science, Krakow, Poland.}
	\AddExternalInstitute{d}{Warsaw University of Technology, Warsaw, Poland.}
	\AddExternalInstitute{e}{Ioffe Physical - Technical Institute of Russian Academy of Sciences, St.~Petersburg, Russian Federation.}
	\AddExternalInstitute{f}{SLAC National Accelerator Laboratory, Stanford CA, USA.}
	\AddExternalInstitute{\dagger}{Deceased.}
\addtocounter{footnote}{-7}
\newcommand{\thefootnote}{\alph{footnote}} 